\documentclass[fleqn,usenatbib]{mnras}

\usepackage[T1]{fontenc}
\usepackage{ae,aecompl}

\usepackage{graphicx}	% Including figure files
\usepackage{amsmath}	% Advanced maths commands
\usepackage{amssymb}	% Extra maths symbols
\usepackage{CJK}    %!!!!!!!!!!!!!!!!!!!!!!!!!!!!!!!!!!!!!!!!!!!!!!!!!!!!!!!!!!!!!!!!!!!!!!!!!!!!!!!!!!!!!!!!!!!!!!!!!!!! Reinsert this for Chinese characters

\usepackage{color}

\newcommand{\bon}{\color{black}}   % {\color{blue}} 

\title[WD 1145+017 Photometric \& X-Ray Observations]{WD 1145+017: Optical Activity During 2016-2017 and Limits on the X-Ray Flux}

\author[Rappaport et al.]{
S.~Rappaport$^1$,
B.L.~Gary$^2$, 
A.~Vanderburg$^{3,4,5}$, 
S.~Xu$^6$  % (许\CJKfamily{bsmi}偲\CJKfamily{gbsn}艺)$^6$,  %%% (?\CJKfamily{bsmi}?\CJKfamily{gbsn}?)}$^4$,
D.~Pooley$^7$,
\newauthor
and K.~Mukai$^{8,9}$
\newauthor
\\
\noindent
$^{1}$ Department of Physics, and Kavli Institute for Astrophysics and Space Research, M.I.T., Cambridge, MA 02139, USA; sar@mit.edu \\
$^{2}$ Hereford Arizona Observatory, Hereford, AZ 85615; BLGary@umich.edu \\
$^{3}$ Harvard-Smithsonian Center for Astrophysics, 60 Garden Street, Cambridge, MA 02138 USA; avanderburg@cfa.harvard.edu \\
$^{4}$  Department of Astronomy, The University of Texas at Austin, 2515 Speedway, Stop C1400, Austin, TX 78712 \\
$^{5}$  NASA Sagan Fellow \\
$^{6}$ European Southern Observatory, Karl-Schwarzschild-Strasse 2, D-85748 Garching, Germany; sxu@eso.org \\
$^{7}$ Department of Physics and Astronomy, Trinity University, San Antonio, TX; dpooley@trinity.edu \\
$^{8}$ CRESST and X-Ray Astrophysics Laboratory, NASA Goddard Space Flight Center, Greenbelt, MD 20771, USA; Koji.Mukai@nasa.gov \\
$^{9}$Department of Physics, University of Maryland, Baltimore County, 1000 Hilltop Circle, Baltimore, MD 21250, USA
}

\date{}

\pubyear{2016}

\begin{document}
\label{firstpage}
\pagerange{\pageref{firstpage}--\pageref{lastpage}}
\maketitle

\begin{abstract}
WD 1145+017 was observed from 2016 November through 2017 June for the purpose of further characterising the transit behaviour of the dusty debris clouds orbiting this white dwarf.  The optical observations were carried out with a small ground-based telescope run by an amateur astronomer, and covered 53 different nights over the 8-month interval.    We have found that the optical activity has increased to the highest level observed since its discovery with {\em Kepler} K2, with approximately 17\% of the optical flux extinguished per orbit.  The source exhibits some transits with depths of up to 55\% and durations as long as two hours.  The dominant period of the orbiting dust clouds during 2016-2017 is 4.49126 hours.  We present `waterfall' images for the entire 2016-2017 and 2015-2016 observing seasons.  In addition, the white dwarf was observed with the {\em Chandra} X-ray Observatory for 10-ksec on each of four different occasions, separated by about a month each.  The upper limit on the average X-ray flux from WD 1145+017 is $\simeq 5 \times 10^{-15}$ ergs cm$^{-2}$ s$^{-1}$ (unabsorbed over the range 0.1-100 keV), which translates to an upper limit on the X-ray luminosity, $L_x$, of $\simeq 2 \times 10^{28}$ ergs s$^{-1}$.  If $L_x \simeq G M_{\rm wd} \dot M_{\rm acc}/R_{\rm wd}$, where $M_{\rm wd}$ and $R_{\rm wd}$ are the mass and radius of the white dwarf, and $\dot M_{\rm acc}$ is the accretion rate, then $\dot M_{\rm acc} \lesssim 2 \times 10^{11}$ g s$^{-1}$.  This is just consistent with the value of $\dot M$ that is inferred from the level of dust activity.
\end{abstract}

% Select between one and six entries from the list of approved keywords. Could we add ``debris disk''?

%%%%%%%%%%%%%%%%%%%%%%%%%%%%%%%%%%%%%%%%%%%%%%%%%%

%%%%%%%%%%%%%%%%% BODY OF PAPER %%%%%%%%%%%%%%%%%%

\section{Introduction}

WD 1145+017 is a unique white dwarf that has the following four attributes: it (1) exhibits atmospheric pollution via an array of metal lines (Vanderburg et al.~2015, hereafter `V15'; Xu et al.~2016); (2) shows strong evidence for a dusty disk which produces IR emission in excess of the white dwarf's intrinsic emission (V15); (3) exhibits deep transits that are thought to be due to orbiting disintegrating debris (V15; Rappaport et al.~2016, hereafter `R16'; G\"ansicke et al.~2016; Gary et al.~2017, hereafter `G17'); and (4) has broad and variable circumstellar metal absorption lines with widths of up to 300 km s$^{-1}$ (Xu et al.~2016).  A history of the discovery of WD 1145+017, a review of its properties, and some ideas about the orbiting debris are given in the review of Vanderburg \& Rappaport (2017, and references therein).  {\bon Some of the basic photometric and spectroscopic properties of the object are summarised in Table \ref{tbl:mags}.}

A substantial fraction of white dwarfs exhibit atmospheric pollution from such metals as Mg, Al, Si, Ca, Ti, Cr, Mn, Fe, and Ni.  This observation, coupled with the relatively short gravitational settling times, suggest that there must be a nearly continual process of accretion (e.g., Zuckerman et al.~2010; Koester et al.~2014).  A smaller fraction (perhaps $\sim$3\%) of all white dwarfs, and up to 20\% of polluted white dwarfs (Zuckerman et al.~1987), are also found to have NIR signatures of dusty disks orbiting them (Kilic et al.~2006; Farihi et al.~2009; Barber et al.~2012; Rocchetto et al.~2015).  

WD 1145+017 has all these attributes, but what makes it unique are (i) the transits which are presumed to be due to dust clouds in 4.5-5 hour orbits (V15; R16; G\"ansicke et al.~2016; G17), and (ii) the very broad absorption lines ascribed to high-velocity metal gases orbiting perhaps even closer to the white dwarf (Xu et al.~2016; Redfield et al.~2017).  The details of the orbiting debris that produces the dust, in particular the numbers and masses of the bodies, are largely unknown.  However, in broad brush, there were 6 distinct periods found in the K2 discovery observations ranging from 4.5 to 4.9 hours; these were named the `A' through `F' periods.  Periods within $\sim$0.2\% of the A period found in K2, have been seen in ground-based observations ever since the transits were first discovered.  The `B' period, at 4.605 hours, was detected briefly, and convincingly, during the 2015-2016 observing season.  The other periods have not yet been detected from the ground.  However, the transit depths measured with K2 for the `C-F' periods were largely below the level of what can be detected from the ground.  All of the periods detected correspond to Keplerian orbital radii close to $\sim$100 white dwarf radii, or approximately 1 $R_\odot$.

Relatively little is known in detail about the masses of the orbiting bodies.  However, a variety of arguments suggest that there probably are at least a dozen bodies capable of emitting dust clouds that are large enough to be detectable with small telescopes.  The masses of these bodies have been estimated to range between $10^{17}-10^{23}$ grams (V15; R16; Veras et al.~2015; Gurri et al.~2016).  The broad metal absorption lines may be due to gas composed of a variety of metals orbiting close in to the white dwarf, perhaps within $\sim$10 $R_{\rm wd}$ (Xu et al.~2016; Redfield et al.~2017; Vanderburg \& Rappaport 2017). 

In this paper we discuss the results of optical monitoring of WD 1145+017 covering 8 months of the 2016-2017 observing season. We also present the results of four {\em Chandra} X-ray observations of 10 ksec each, but spanning an interval of 20 weeks.  In Sects.~\ref{sec:obser} and \ref{sec:chandra} we describe how the optical and X-ray observations were acquired.  The results of the optical observations are presented in \ref{sec:optical}, including a number of illustrative lightcurves (Sect.~\ref{sec:LCs}), a new way to visualise `waterfall' diagrams (Sect.~\ref{sec:WF}), a search for periodicities (Sect.~\ref{sec:periods}), and an overview of the photometric dip activity since its discovery (Sect.~\ref{sec:activity}).  The  {\em Chandra} X-ray observations led to only an upper limit on the flux which is evaluated and discussed in Sect.~\ref{sec:Xray}.  An interpretation of what this limit on the X-ray flux implies for the accretion rate of debris is explored in Sect.~\ref{sec:otherbands}.  Our current understanding of WD 1145+017 is discussed more broadly in Sect.~\ref{sec:discuss}.  We summarize our results and draw some final conclusions in Sect.~\ref{sec:conclusion}.

\begin{table}
\centering
\caption{{\bon Photometric and Spectral Properties of WD 1145+017}}
\begin{tabular}{lc}
\hline
\hline
Parameter &
WD 11145+017 \\
\hline
RA (J2000) & 11:48:33.627   \\  
Dec (J2000) &  +01:28:59.41  \\  
Spectral Type & DBZA \\
$K_p$$^a$ & $17.29$ \\
%u$^a$ & 16.97  \\  % SDSS
g$^b$ &  $17.00 \pm 0.01$ \\  % SDSS
%r$^a$ & 17.38  \\  % SDSS
%i$^a$ & 17.60 \\ % SDSS
%z$^a$ & 17.84  \\ % SDSS
%Y$^b$ & 17.43 \\ % UKIDSS
J$^c$ & $17.50 \pm 0.03$ \\ % UKIDSS
%H$^c$ & 17.53 \\
K$^c$ & $17.40 \pm 0.08$ \\ % UKIDSS
W1$^d$ & $17.02 \pm 0.16$ \\ % WISE
W2$^d$ & $16.51 \pm 0.35$ \\ % WISE
$T_{\rm eff}$$^e$ (K) & $15,900 \pm 500$ \\
$\log \, g$$^e$ (cgs) & $8.0 \pm 0.2$ \\
$R_{\rm wd}^e ~(R_\oplus)$ & $1.40 \pm 0.18$ \\
$M_{\rm wd}^e ~(M_\odot)$ & $0.59 \pm 0.12$ \\
Cooling Age$^e$ (Myr) & $175 \pm 75$ \\
Distance (pc)$^f$ & $174 \pm 25$  \\   
$\mu_\alpha$ (mas ~${\rm yr}^{-1}$)$^f$ & $-43.3 \pm 4.9$  \\ 
$\mu_\delta$ (mas ~${\rm yr}^{-1}$)$^f$ &  $-7.0 \pm 4.9$  \\ 
orbital period$^g$ (hrs) & 4.49126 \\
orbital radius$^g$ ($R_\odot$) & 1.16 \\
\hline
\label{tbl:mags}
\end{tabular}

{\bon {\bf Notes.} (a) \url{https://archive.stsci.edu/k2/epic/search.php}.  (b) Taken from the SDSS image (Ahn et al.~2012).  (c) UKIDSS magnitudes; Lawrence et al.~(2007).  (d) WISE point source catalog (Cutri et al.~2013). (e) Vanderburg et al.~(2015); \url{http://dev.montrealwhitedwarfdatabase.org/evolution.html}. (f) From UCAC4 (Zacharias et al.~2013); Smart \& Nicastro (2014). (g) The primary periodicity observed in this work.}
\end{table}

\section{Ground-Based Optical Monitoring}
\label{sec:obser}

\begin{figure*}
\begin{center}
\includegraphics[width=0.65 \textwidth]{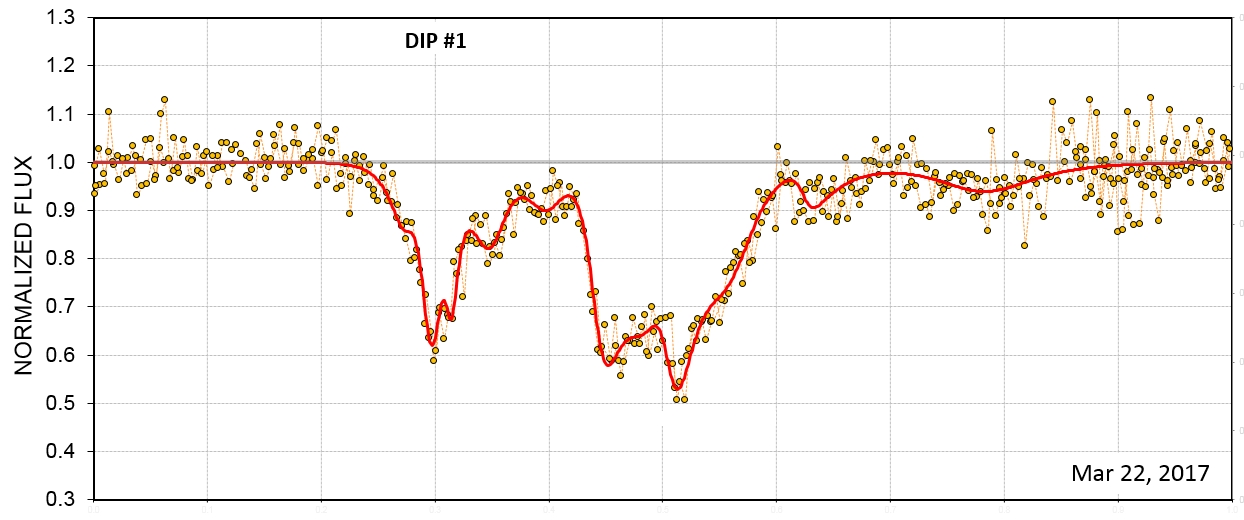}
\includegraphics[width=0.65 \textwidth]{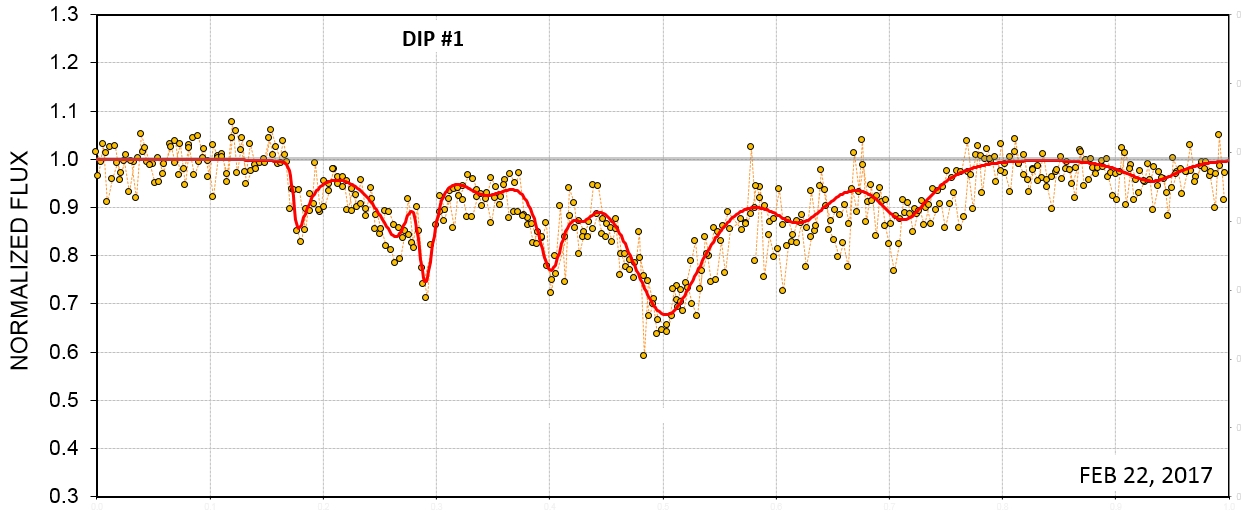} \hglue0.02cm
\includegraphics[width=0.65 \textwidth]{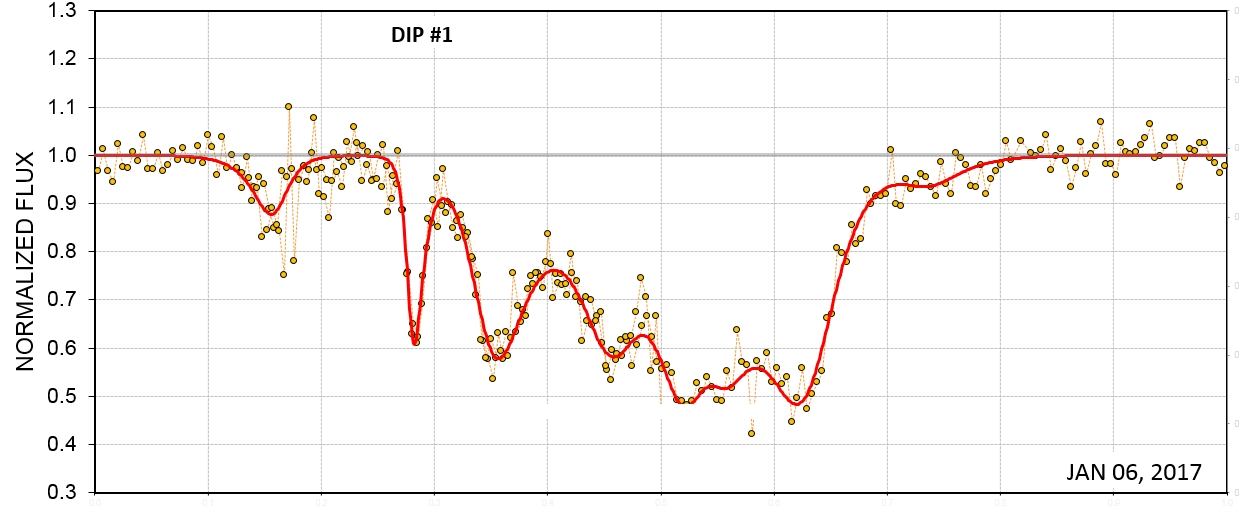} \hglue0.22cm
\includegraphics[width=0.66 \textwidth]{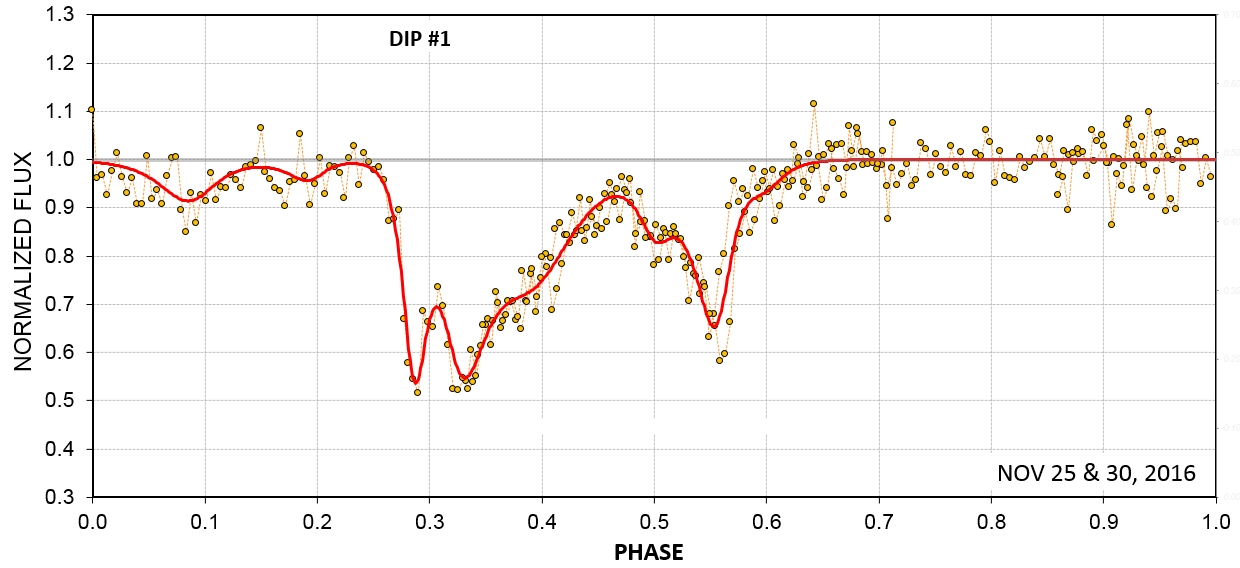} 
\caption{Four lightcurves of WD 1145+017 showing normalised flux for each image (small circles) and a fitted model (red curve; see text).  Note the deep (up to 50\%) and very long dips which were not present during the last three observing seasons.  The lightcurves are phased to a period of 4.49126 hours and epoch of phase zero at BJD 2457687.7335.}
\label{fig:LCs} % Figure 1
\end{center}
\end{figure*}

We report on 53 observing sessions of WD 1145+017 during the 8-month interval 2016 October 25 through 2017 June 18. The median interval between observing sessions was four days.  

All observations were made with the Hereford Arizona Observatory (HAO, Minor Planet Center site code of G95). This is a private observatory in Hereford, Arizona, consisting of a 14-inch Meade LX200 GPS telescope and a Santa Barbara Instrument Group (SBIG) ST-10XME CCD camera.  {\bon All observations were unfiltered; the system response exceeded 10\% of maximum from 400 to 900 nm.} The telescope, camera, filter wheel, autoguiding systems and dome azimuth are controlled from a residence office via buried 100-foot cables. A commercial program, MaxIm DL v5.2, is used for observatory control as well as later image processing. Calibration images (bias, dark, and flat) are updated at appropriate intervals, with emphasis on using the same CCD camera control temperature for dark frames. Autoguiding was usually performed using the ST-10XME second chip (meant only for autoguiding) in order to preserve the star field's pixel location throughout each observing session. Once an observing session is set up and operating in a stable manner the observatory is allowed to function for the rest of the night unattended; automatic shut-down of hardware is accomplished when specified conditions are met (elevation $< 15^\circ$, astronomical twilight, etc.). 

The image processing, using MaxIm DL, is maintained as close as possible to the same procedure for all observing sessions throughout the entire observing season. All images are calibrated, and then star aligned. An artificial star is placed in the upper-left corner of each image for the purpose of converting the photometry tool's magnitude measurements to flux (since MaxIm DL v5.2 does not permit saving flux information). The photometry tool is used to specify WD 1145+017 as the ``target'', the artificial star is specified as a reference star, and 25 nearby stars are specified as ``check stars''.  The photometry target circle radius is set to 4 pixels ($7''$), the gap annulus width is set to 3 pixels and the sky background annulus width is set to 12 pixels; photometry readings are saved as text files. The photometry signal circle radius is also changed to 3 and 5 pixels, and these two additional photometry data files are saved. 

The rationale for the preceding steps, and those to be presented next, are described in the book {\em Exoplanet Observing for Amateurs} (Gary 2014). The procedures were developed during 15 years of exoplanet observing and the similarities with the WD 1145+017 light curve generation made it an obvious choice for the present work. 

Each of the text files with photometry readings is imported to a spreadsheet specifically meant for use with WD 1145+017. For each of the three photometry target circle radius readings a calculation is made of the RMS(t) variations using a neighbour difference method.   A weighted average of the three photometry readings is calculated; this achieves most of the optimisation that would be accomplished if one photometry reading were made with a dynamic photometry aperture size, so the resultant set of weighted-average magnitudes are less affected by atmospheric seeing changes to the point-spread-function (FWHM size) during the observing session. The user manually solves for atmospheric extinction, including a temporal trend term, with the aid of graphs. Departures from the extinction model reveal when clouds were present, as well as other anomalies (such as large seeing degradations). 

A special method is used to automatically identify artefacts in the 25 stars (identified within MaxIm DL as ``check''), such as cosmic ray hits, and these specific star readings are disregarded in a process that produces total flux for all (accepted) check stars; the final total flux vs. UT is treated like a ``reference star''.  Graphical displays provide a means for quickly identifying stars that are unexpectedly variable (or saturated, and therefore not suitable for use); the user can toggle ``use/don't-use'' spreadsheet switches for each of the 25 stars. The use of ``sum of flux'' for all accepted stars for ``reference'' is superior to averaging magnitudes, as is done by standard differential photometry. 

The resulting WD 1145+017 magnitudes are assigned standard errors (`SE') uncertainties by calculating RMS(t) for several of the 25 stars used for reference, and a model for RMS vs. instrument magnitude is used for predicting the WD 1145+017 SE. A neighbour-difference method (Gary 2014) is used for calculating RMS(t) for the reference stars.  

The spreadsheet analysis procedure is not meant for automatic implementation, as it requires user judgement at many steps. However, most decisions are guided by objective calculations to minimise subjective bias while assuring data quality.  Because of this extra attention, and because the ST-10XME CCD has high QE (87\%), it has been estimated that the HAO 14-inch telescope system performs like a typical 17-inch telescope. 

\begin{figure*}
\begin{center}
\includegraphics[width=0.75 \textwidth]{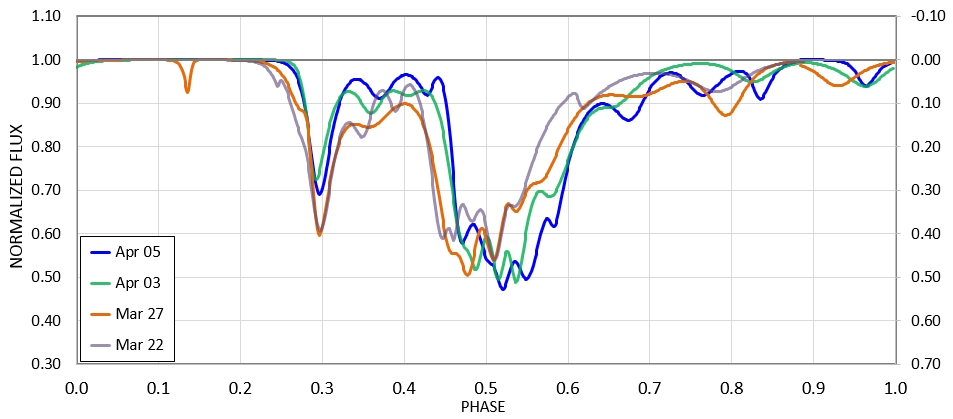} 
\includegraphics[width=0.75 \textwidth]{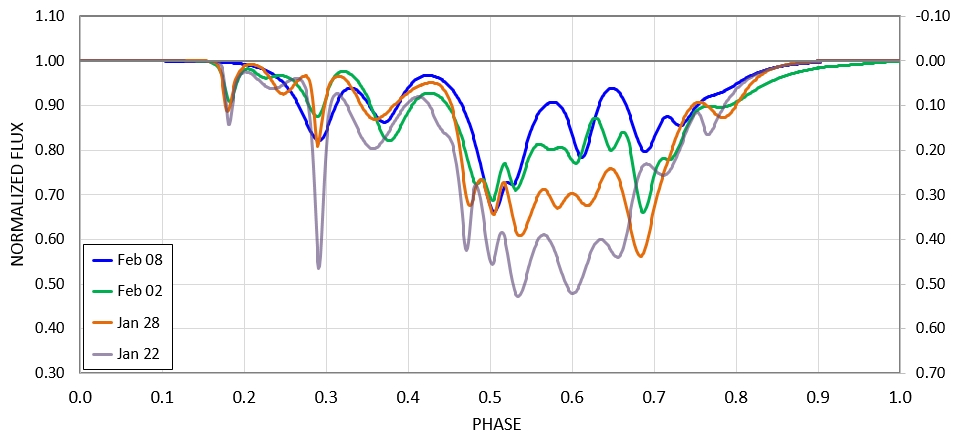} 
\caption{Superpositions of fits to four different light curves covering the interval JD 245,7834-48 (top panel) and JD 245,7775-92 (bottom panel).  The fits are each to the lightcurve from a single night.  They are presented in lieu of the data themselves to promote clarity and minimise confusion.  The same period and epoch are used here as in Fig.~\ref{fig:LCs}.}
\label{fig:superposed} % Figure 2
\end{center}
\end{figure*}

\section{{\em Chandra} X-Ray Observations}
\label{sec:chandra}

Because of the possibility of observing X-rays from the inferred accretion of material onto the white dwarf in WD 1145+017, we proposed {\em Chandra} observations in Cycle 18.  We requested exposures of 10 ksec to be carried out on four different occasions to be separated by at least one month.  We did this because (i) it was difficult to predict in advance when WD 1145+017 would be most active in terms of exhibiting dust extinctions, and (ii) there is an unknown time delay between the release of dust in the system, and the subsequent accretion of the sublimated gas from that dust.  Thus, we elected to search for X-ray emission over a range of different times.

The {\em Chandra} proposal was accepted and the observations were carried out during February, April, May, and July of 2017.  Each observation was 10 ksec in duration, for a total exposure of 40 ksec. {\bon The dates of the observations are summarised in Table \ref{tbl:Chandra}}.

{\bon The observations were made with ACIS-S (of the Advanced CCD Imaging Spectrometer) using the back-illuminated S3 chip in imaging (i.e., non-grating) mode.  We utilized the timed exposure (TE) and very faint (VF) source modes, and with no subarrays.  The on-board energy filtering range was 0.24 to 12 keV. }

\begin{table}
\centering
\caption{{\em Chandra} Observations of WD 1145+017}
\begin{tabular}{llcc}
\hline
\hline
Date (JD) &  Date (UT) & {\bon Obs.~ID} & Exposure (ksec)  \\
\hline
2457802.3 & 2017 Feb.~17 & {\bon 18917} & 10.4 \\ 
2457850.1 & 2017 April 6 & {\bon 18918} & 9.9 \\ 
2457902.0 & 2017 May 28 & {\bon 18919} & 9.9 \\ 
2457939.1 & 2017 July 4 & {\bon 18920} & 9.9 \\
\hline
Total & $\sim$20 weeks & ... & 40.2 \\
\hline
\hline
\end{tabular}
\label{tbl:Chandra}

\scriptsize{} 

\end{table}

\section{Results of the Optical Studies}
\label{sec:optical}

\subsection{Light Curves}
\label{sec:LCs}

During this past observing season (2016-2017), the dips in the optical flux from WD 1145+017 were no deeper than during the previous season, but they were more numerous, often resulting in effectively very long (1.5-2 hour) transits.  In net, the dips produced the highest ``activity'' level seen since the source's discovery as a white dwarf undergoing periodic transits (V15).  As much as 17\% of the average flux during a 4.5-hr period was missing in January (see Fig.~\ref{fig:activity}).

In Figure \ref{fig:LCs} we show four illustrative lightcurves recorded in November, January, February, and March of this past 2016-2017 observing season.  Except for the November observation, the lightcurve covers a full orbital cycle recorded during a single night, and is plotted against orbital phase using a common period of 4.49126 hours (see also Croll et al.~2017), and an epoch of JD 2457687.7335.  In all cases, there is a portion of the flux dip that lasts for at least 0.3 orbital cycles, and in two cases for nearly half an orbital cycle.  The dips in flux range in depth, at their deepest points, up to 55\%, but were never seen to go deeper than 55\% at any time during this season's observations.  The solid curves superposed on the data are fits to sums of asymmetric hyperbolic secants (`AHS') which we have discussed in our previous work (R16; G17).  They are meant to (i) guide the eye, (ii) assess quantitatively the area under the dip, and (iii) track phase drifts of different persistent features.  

In order to better visualise how the lightcurves evolve over timescales of days to weeks, we show some illustrative examples in Fig.~\ref{fig:superposed} of evolving lightcurves during the intervals 2017 Jan.~22 to Feb.~8, and March 22 to April 5.  In each case we show the AHS fits to the lightcurves on four different days which span an interval of about two weeks.  These curves are all phased to the same period (4.49126 hours) using approximately the same epoch of phase zero as that in Fig.~\ref{fig:LCs}.  Figure \ref{fig:superposed} serves to show how, on the one hand, the dip features are quite repeatable from day to day, while on the other hand, they do clearly evolve on timescales of weeks.

\begin{figure*}
\begin{center}
\includegraphics[width=0.45 \textwidth]{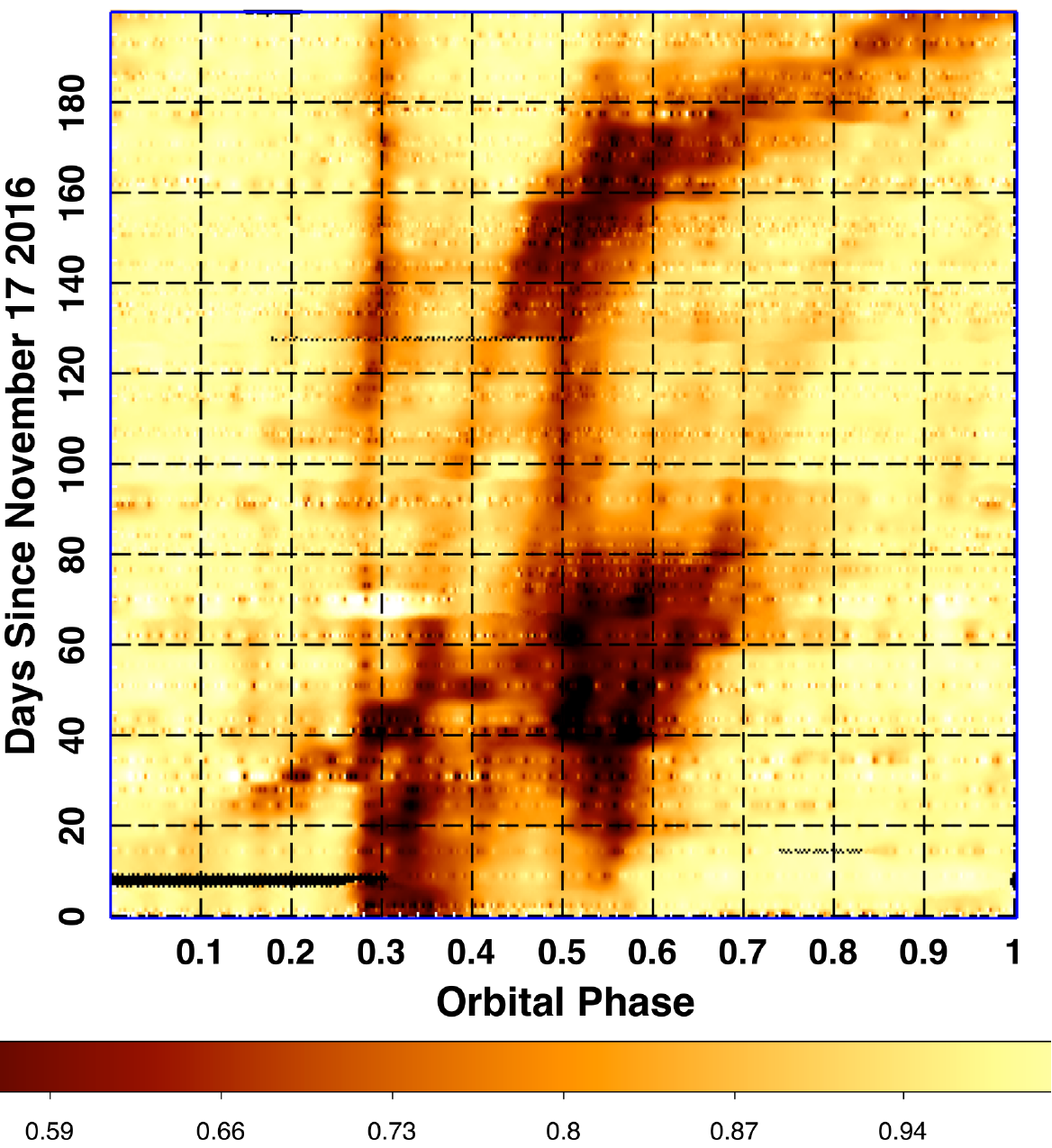} \hglue0.2cm \vspace{-0.2cm}
\includegraphics[width=0.45 \textwidth]{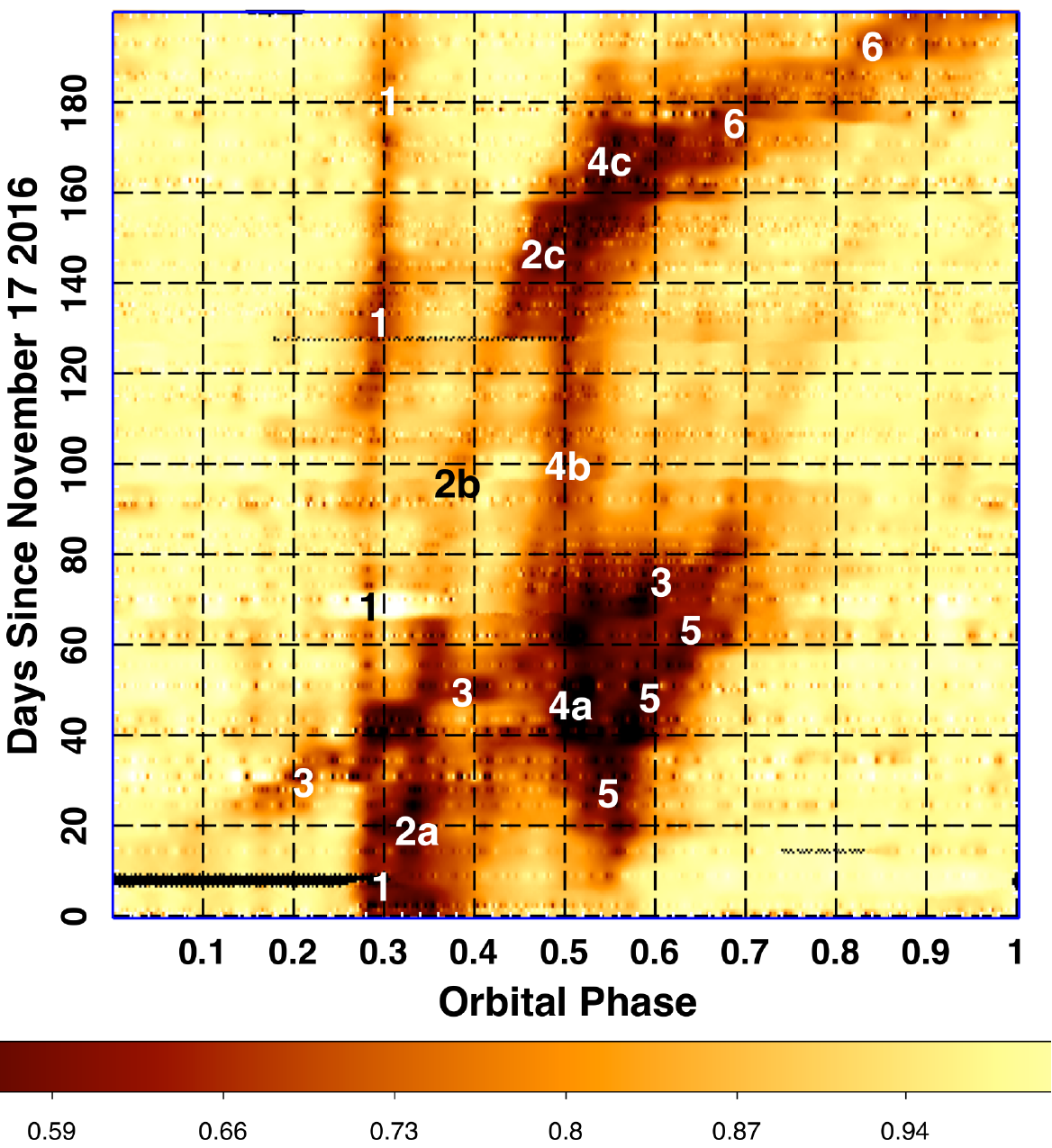}
\caption{WD 1145+017 waterfall diagrams for the 2016-2017 season.  These were produced from the lightcurves obtained on 53 individual nights, and then interpolated to fill in the regions where data are missing (see text for a discussion of the interpolation method).  The right-hand panel is the same figure, but with various features numbered for discussion in the text.  The color-coding represents the normalised source brightness, with yellow-orange regions being near full intensity, while the darkest regions have only $\sim$50\% of the full brightness.  The period used to phase both diagrams is 4.49126 hours, and the epoch of phase zero is taken, somewhat arbitrarily, to be BJD 2457687.7335.}
\label{fig:waterfall} % Figure 3
\end{center}
\end{figure*}

\subsection{Waterfall Diagrams}
\label{sec:WF}

Perhaps the best way to obtain an overview of the dip activity in WD 1145+017 over an observing season is to study a ``waterfall'' type diagram.  In previous work, we have produced waterfall diagrams by (i) stacking traces of the flux vs.~orbital phase in order of observation date (see, e.g., Fig.~1 of Rappaport et al.~2016; also Fig.~1 of G\"ansicke et al.~2016), and (ii) reducing each fitted dip to a bar whose length is the dip duration, and thickness is the dip depth (e.g., Fig.~6 of R16; Fig.~4 of G17).

In this work we attempt a new approach which we believe presents a more coherent picture of the source activity.  The waterfall diagram nominally indicates the orbital phase along the x-axis and observation date along the y-axis.  We have constructed a waterfall-diagram `image' of 400 $\times$ 400 pixels where the intensity in each pixel represents the measured flux at a specific moment.  Each pixel in the x direction represents 40 sec in orbital phase (comparable to the cadence of the observations) while each pixel in the y direction corresponds to a half day on the calendar.  

The results for the 2016-2017 season are shown in Fig.~\ref{fig:waterfall}.  The challenge was to fill in the quite substantial gaps where no observations were conducted.  As mentioned above, the median gap between observations was about 4 nights.  In addition some of the observations early and late in the season suffered from shortened windows of target visibility.  We handled the data gaps as follows.  The flux from each observation was placed into the appropriate $\{x,y\}$ bin according to the phase of the 4.5-hr period and the date of the observation.  We then applied a simple algorithm to fill in the gaps, as follows.  For each point in the image, if there exists a data point, we leave it as is.  If a pixel is initially blank, we draw a circle around that point which is 10 pixels in radius (corresponding to 7 minutes in phase, and 5 days on the calendar), and take a distance weighted average of all the points within which there are data.  The weighting was done according to $d^{-2}$, where $d$ is the distance between the point being ``filled'' and the data points (in units of pixels).  Specifically, the flux used to fill the blank bin is given by
\begin{equation}
F_n = \sum_{m \ne n} \frac{F_{mn} d_{mn}^{-2}}{d_{mn}^{-2}}
\end{equation}
where $m$ is any other point within 10 pixels of $n$.  

\begin{figure}
\begin{center}
\includegraphics[width=0.96\columnwidth]{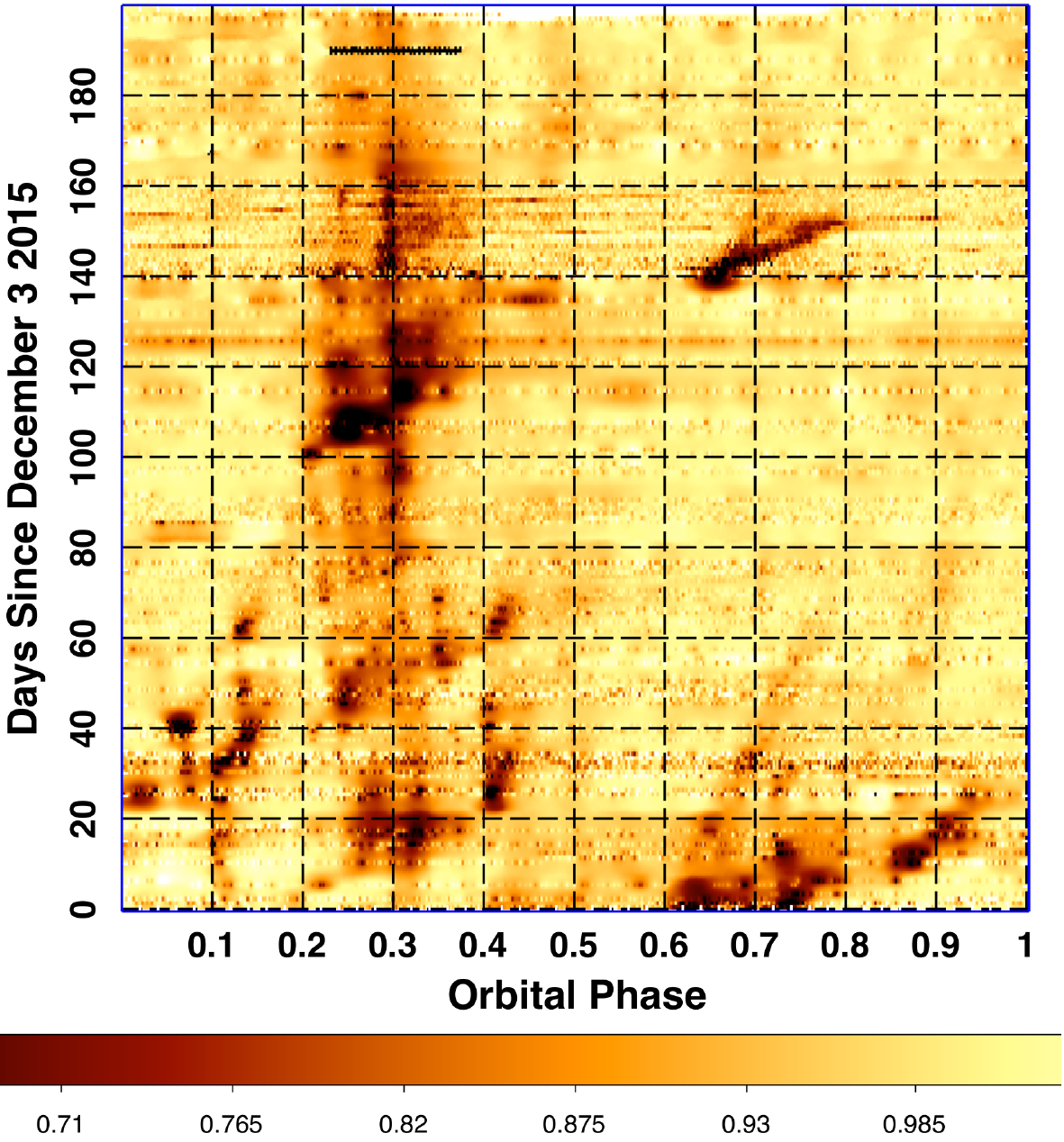} 
\caption{WD 1145+017 waterfall diagrams for the 2015-2016 season.  Other descriptors are the same as for Fig.~\ref{fig:waterfall}.  The period used to phase both diagrams is 4.49126 hours, and the epoch of phase zero is taken, somewhat arbitrarily, to be JD 2457687.7335. }   %  8083
\label{fig:WFprevious}  % Figure 4
\end{center}
\end{figure}

What we see from the resultant waterfall plot based on data from the 2016-2017 season is that many of the features have a period close to 4.49126 hours.  That this is the case can be understood from the fact that features in the image with this period tend to run vertically.  Just to indicate the sensitivity of such a plot to periods that may differ from the fold period, we point out that a feature lying at a 45$^\circ$ angle represents a period difference of $\Delta P/P \simeq +0.001$.  In the right-hand panel of Fig.~\ref{fig:waterfall} we have numbered 6 of the dip features which seem to persist for intervals of days to months.  We emphasise, however, that via these diagrams alone there is not a robust way to uniquely identify the separate features, and the ones we have numbered are subjective.

Here we describe more specifically the various numbered dip features in Fig.~\ref{fig:waterfall}.  
Feature 1 is mostly a straight vertical line, but with some kinks or bends.  The smallest discernible curvatures correspond to $P/\dot P \simeq 2000$ years.
Feature 2 exhibits a large degree of curvature---corresponding $P/\dot P \simeq 400$ years.
Feature 3 is `highly' sloped with a period of 4.4953 hours vs.~the 4.49126 hours used to make the fold. 
Feature 4 is a new dip that also starts around day 40 and ends around day 180.  Initially at least it seems to have the same period as Feature 1, but simply appears at a different orbital phase. 
Feature 5 seems to be a new dip feature that appears around day 20 of the plot and ends about two months later.
Feature 6 starts at about day 170 and persists until the end of the observations.  It may have a period that is similar to that of Feature 3.

One thing that should be kept in mind about these features is that they all have orbital periods within a range of 0.1\%.  The longest period, 4.4953 hr, is closer to the K2 `A' period at 4.4989 hr.

We next went back to the photometric data that we had collected during the previous observing season (2015-2016) and utilised the same algorithm to construct the corresponding waterfall plot.  The results are shown in Fig.~\ref{fig:WFprevious}.  The fold period and epoch of the fold are the same as used to produce Fig.~\ref{fig:waterfall}.  Many of these features were discussed extensively in G17, so we will point out only the most salient features.  First there is a prominent vertical stripe which appears to be closely connected to Feature 1 in Fig.~\ref{fig:waterfall}; it may very well be the same underlying body, in a 4.49126 hour period, that accounts for both these features in Figs.~\ref{fig:waterfall} and \ref{fig:WFprevious}.  The second set of prominent features are the three sets of dips that run at roughly an angle of 30$^\circ$ from horizontal.  These correspond to a period of 4.4990 hours (very close to the K2 A period), but with a wide range of orbital phases for the three sets of dips.  Finally, there is a feature that seems to begin abruptly at day 140 and phase 0.65.  This is in fact the event denoted as `G6420' in G17, and shown in greater detail in Fig.~6 of G17.

\begin{figure}
\begin{center}
\includegraphics[width=0.98\columnwidth]{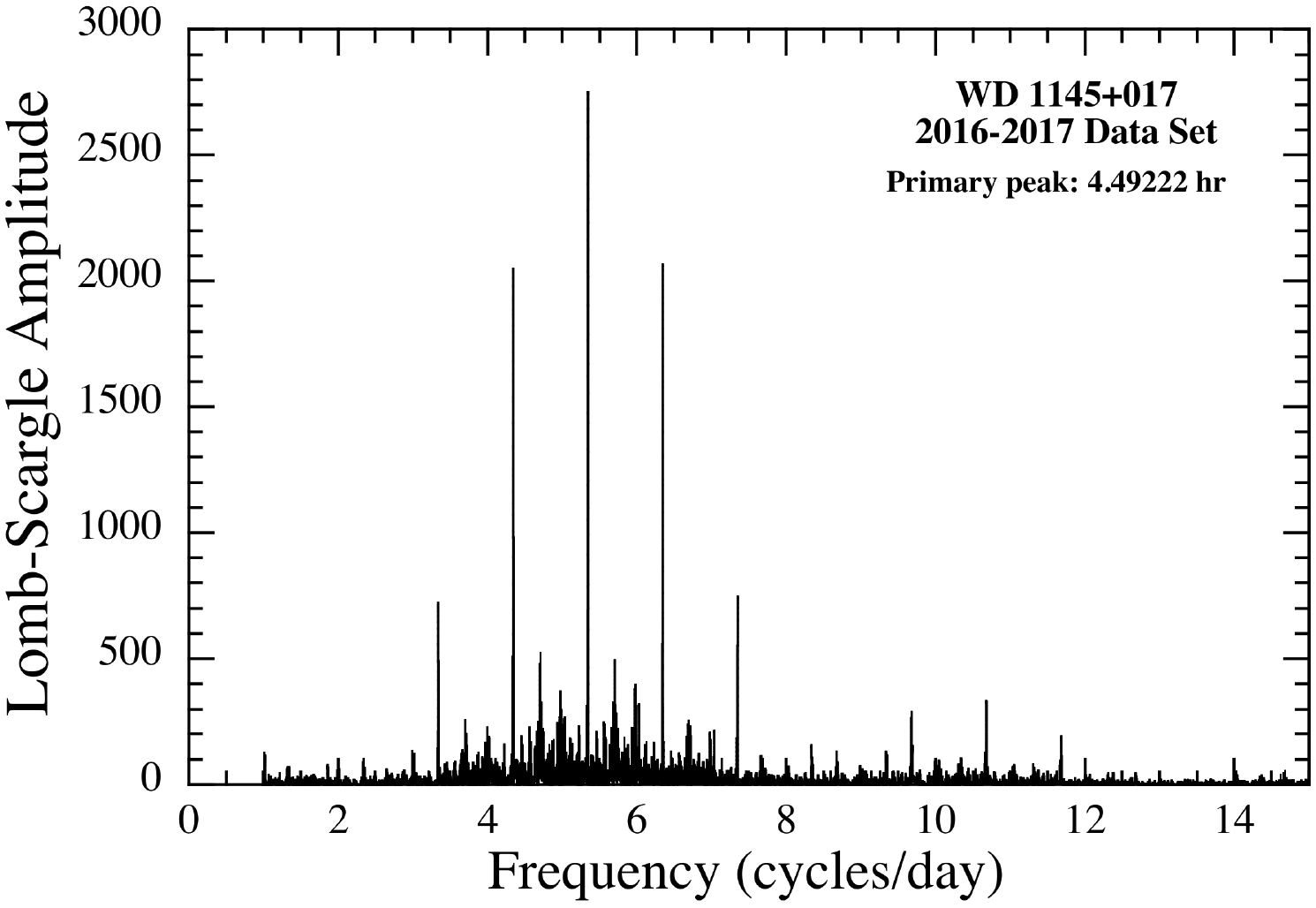} \hglue0.09cm
\includegraphics[width=0.98\columnwidth]{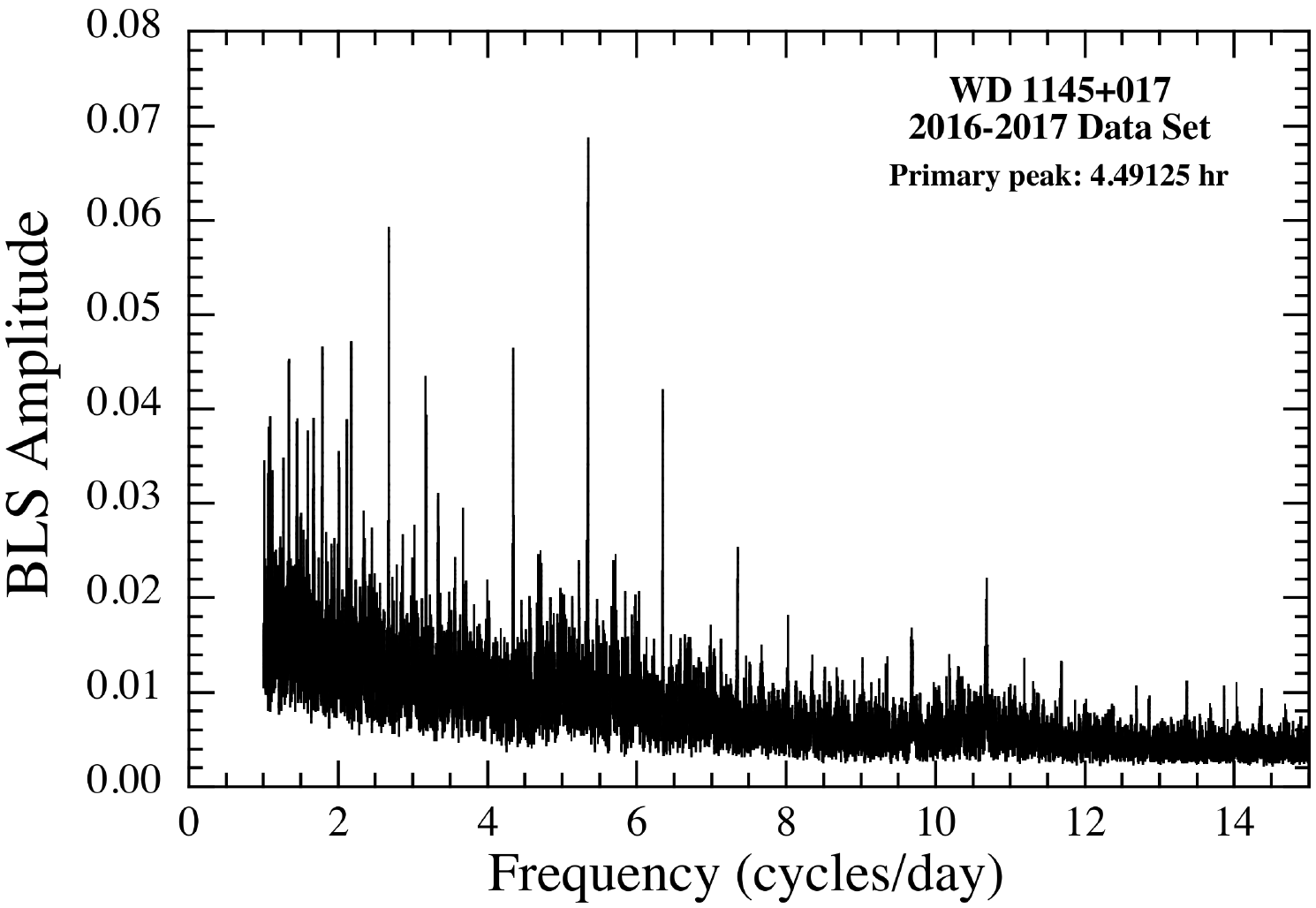} 
\caption{Periodograms of the WD 1145+017 optical flux data for the 2016-2017 observing season.  Top panel: Lomb-Scargle transform.  Bottom panel: Box Least Squares transform.  The principal periods detected are $4.4922 \pm 0.0003$ hours in the L-S transform and $4.4912 \pm 0.0004$ hours in the L-S transform.  Most of the other significant peaks are 1-day aliases of the principal period, or its higher harmonics. }
\label{fig:periodogram}  % Figure 5
\end{center}
\end{figure}

\subsection{Search for Periodicities}
\label{sec:periods}

In order to search for periodicities that might be persistent, but of too low an amplitude to show up in the waterfall diagram (Fig.~\ref{fig:waterfall}), we carried out both a Lomb-Scargle (Lomb 1976; Scargle 1982) as well as a Box Least Squares periodogram (Kov{\'a}cs et al.~2002) search.  In each case, we used all of the ground-based optical photometry data taken during the 2016-2017 season, subtracted the mean flux, and then subjected the resultant data set to these transforms.  

The results are shown in Fig.~\ref{fig:periodogram}.  The top panel shows the Lomb-Scargle transform.  The centroid of the most prominent peak is at 4.4922 hours with a nominal width of 0.004 hours.  Given that this peak is detected at about the 15-$\sigma$ level, we estimate the uncertainty in the period at $\sim$0.0003 hours. All the other significant peaks in the L-S transform are either aliases of the 1-day observational window or the higher harmonics of the main peak and its 1-day aliases.  The lower panel in Fig.~\ref{fig:periodogram} shows the results of the BLS transform.  The highest peak occurs at 4.4912 hours with an uncertainty of $\sim$0.0004 hours.  The two periods are roughly consistent to within the statistical uncertainties.  

It is worth noting that if any periodicities have an amplitude that is an order of magnitude lower than that of the prominent 4.4913 hour period reported here, it would not show up in either the L-S or BLS transforms as a significant signal.  Thus, we can conclude that the K2 `B-F' periods, with the amplitudes found in the original K2 data ($\lesssim$1/2\%) would not have been detected here. In this regard, we note that the `B' period did become much more prominent for a couple of weeks in 2016 April-May (see G17).  However, it was identified because individual transits were seen by eye, and was not found via a periodicity search.  

\begin{figure}
\begin{center}
\includegraphics[width=0.99 \columnwidth]{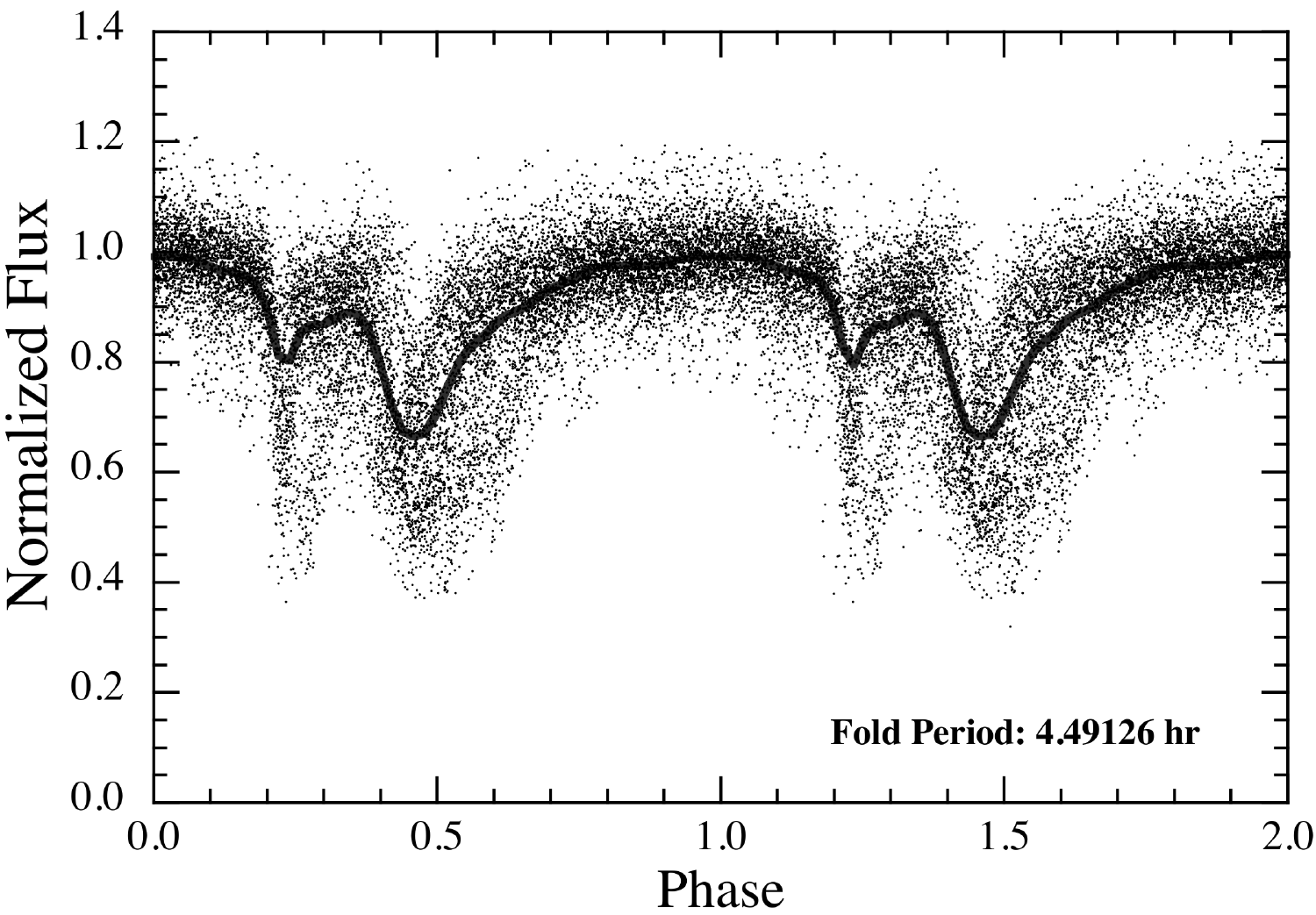} 
\caption{Flux data from WD 1145+017 folded about a period of 4.49126 hours.  The black dots are the individual data points, while the heavy black curve represents a smoothed profile.}
\label{fig:fold} % Figure 6
\end{center}
\end{figure}

Finally, in regard to the principal period found in this data set, we show in Fig.~\ref{fig:fold} a fold of all the 2016-2017 photometry data about a period of 4.49126 hours.  

\subsection{Activity Level of WD 1145+017}
\label{sec:activity}

We have collected all of the data available to us on the source `activity' since it was discovered (V15).  The `activity' level is defined formally by Eqn.~(1) of Gary et al.~(2017), but is essentially just the integrated dip depth over one orbital cycle.  We show an updated version of the activity level diagram, including this entire current observing season, in Fig.~\ref{fig:activity}.  Note that, at times during this past season, the activity level has been two orders of magnitude greater than during the K2 discovery observations.  On average, the source has been twice as active this season compared to last season.

\section{Inferences from Limits on X-Ray Flux}
\label{sec:Xray}

\subsection{X-Ray Image}
\label{sec:Ximage}

\begin{figure*}
\begin{center}
\includegraphics[width=0.83 \textwidth]{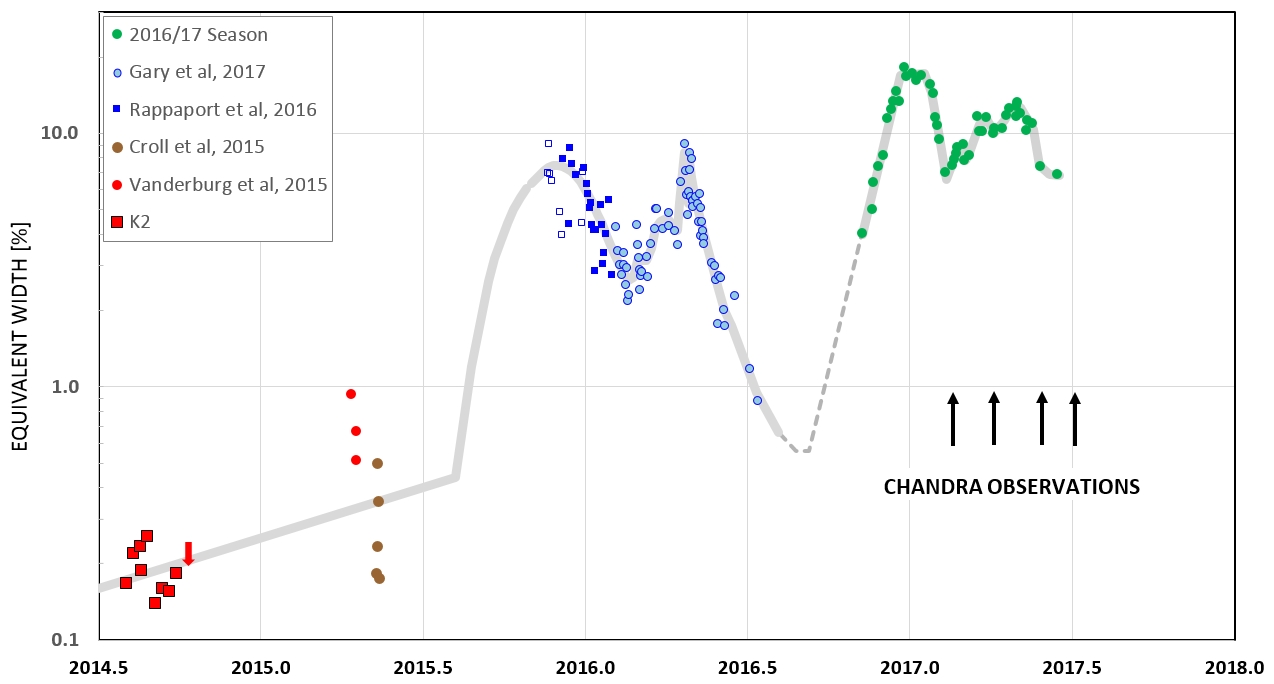} 
\caption{The `activity' level of WD 1145+017 from the time of its discovery with K2.  `Activity' is defined as the fraction of the flux, averaged over one 4.5-hour period, that is extinguished by dust. The activity this season was the highest on record. The times of the {\em Chandra} observations are marked by the four arrows. }
\label{fig:activity} % Figure 7
\end{center}
\end{figure*}

	As discussed in Sect.~\ref{sec:chandra} there were four 10-ksec {\em Chandra} exposures of WD 1145+017 during the period 2017 February 17 to July 4.  The times of the four {\em Chandra} exposures, relative to the optical activity, are shown as arrows on the activity plot (Fig.~\ref{fig:activity}).  As is obvious from the plot, three of the four exposures were taken at among the highest dust-activity levels known since the source was discovered.  The final {\em Chandra} observation was carried out just a month after the end of this season's ground-based observations.  However, based on the previous source activity evolution, it seems reasonable to assume that the level of dust production was still quite substantial.  
	
\begin{figure*}
\begin{center}
\includegraphics[width=0.45 \textwidth]{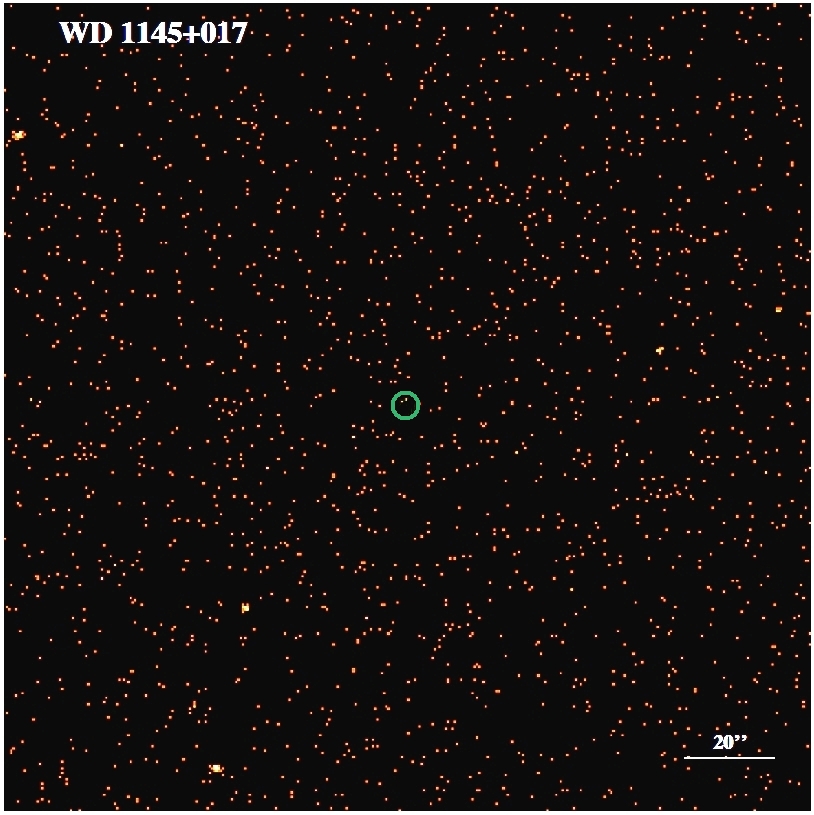} \hglue0.2cm
\includegraphics[width=0.45 \textwidth]{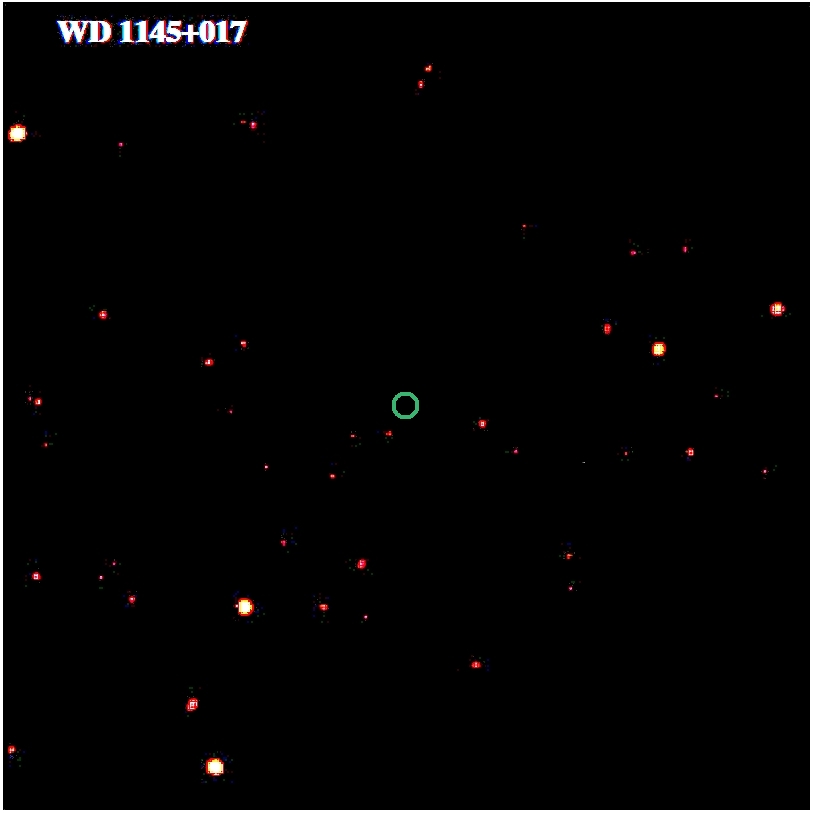} 
\caption{{\em Chandra} X-ray image of the $3' \times 3'$ region centred on WD 1145+017. The exposure is 40 ksec, and the energy band is 0.5-6 keV.  The raw image is shown in the left panel where all the individual detected photons are displayed.  In the panel on the right, the X-ray flux has been smoothed with a kernel of $1.5''$ radius.  The location of WD 1145+017 is indicated by the green circle of 3$''$ radius.  No significant X-ray detection is made of WD 1145+017, with a 95\% confidence upper limit of $3 \times 10^{-15}$ ergs cm$^{-2}$ s$^{-1}$.}
\label{fig:chandra} % Figure 8
\end{center}
\end{figure*}

	The X-ray image produced from the sum of the four separate exposures is shown in Fig.~\ref{fig:chandra}; the image covers a $3' \times 3'$ region.  The left panel shows the raw X-ray image with all the individual detected X-rays displayed.  The location of WD 1145+017 is marked with a small green circle.  The right-hand panel shows the same image smoothed with a Gaussian kernel of $1.5''$ radius.  While there are about a half dozen weak background X-ray sources in the image, there is actually only a single detected X-ray photon within $1''$ of the target location. 

\begin{table*}
\centering
\caption{{\em Chandra} Flux/Luminosity Limits on WD 1145+017}
\begin{tabular}{lcccccc}
\hline
\hline
Model$^a$ &  Parameter & Absorbed Flux Limits$^b$  & \multicolumn{2}{c}{Unabsorbed X-Ray Flux Limits$^b$}   & $\log L_x$$^c$  &  $\dot M$$^d$ \\
                  &                  & 0.5-6 keV &  0.5-6 keV  & 0.1-100 keV  & 0.1-100 keV &      \\
\hline
APEC  &    1 keV  & 1.24 &  1.34  &  1.77  &   27.81  &   0.73 \\
APEC  &   2 keV   & 1.23   & 1.30    & 1.74  &   27.80  &   0.71 \\
APEC  &  5 keV  & 1.44   & 1.50   & 2.34  &   27.93   &  0.96 \\
APEC &   10 keV  &  1.53  &  1.59   &  3.16  &   28.06   &  1.30 \\
APEC  &  20 keV  & 1.60  &  1.65   &  4.65   &  28.23   &  1.91 \\
APEC  &  40 keV &  1.63  &  1.67   & 7.02   &  28.41   &  2.88 \\
APEC  &  64 keV  & 1.64   & 1.69   & 9.02   &  28.52   &  3.71 \\
Bremss. &  20 keV &  1.56  &  1.61  &  4.80  &   28.25  &   1.97 \\
Bremss. & 40 keV   & 1.60   & 1.65  &  7.06  &   28.41  &   2.90 \\
Bremss. & 64 keV  & 1.62   & 1.66  &  8.90  &   28.51   &  3.66 \\
Bremss. & 80 keV  &  1.62  &  1.67  &  9.74  &   28.55  &   4.00 \\
BB   &    1 keV &  1.71  &  1.73 & 2.02   &  27.87   &  0.83 \\
BB    &   2 keV  & 2.10  &  2.11   & 5.38  &   28.29   &  2.21 \\
PL   &    $\alpha = 2$    &   1.46  &  1.54   & 4.29  &  28.20  &   1.76 \\
PL   &    $\alpha = 1$    &   1.71  &  1.74  & 31.7   &  29.06   & 13.0 \\
\hline
Median$^e$ & ... & 1.60 & 1.65 & 4.80 & 28.2 & 2.0 \\ 
\hline
\hline
\end{tabular}
\label{tbl:xray}

\scriptsize{(a) Models: {\em APEC} plasma; Bremss.~= thermal bremsstrahlung; BB = blackbody; PL = power law.  (b) The units for the absorbed and unabsorbed flux are both $10^{-15}$ ergs cm$^{-2}$ s$^{-1}$.  The unabsorbed flux is given for two different energy ranges as indicated.  (c) Units of $L_x$ are ergs s$^{-1}$; assumes a source distance of 175 pc.  (d) Inferred upper limit on the accretion rate in units of $10^{11}$ grams s$^{-1}$ under the assumption that the accretion luminosity comes out in the {\em Chandra} band (see Sect.~\ref{sec:otherbands}), and that Eqn.~(\ref{eqn:mdot}) holds. (e) Median value of all the column entries.} 

\end{table*}

\subsection{Limits on X-Ray Flux and Luminosity}
\label{sec:Xlimit}

	The fact that only one X-ray is detected in 40 ksec of exposure implies that we are essentially dealing with an upper limit to the flux.  If we hypothesise a mean count rate for the source of 6 photons in 40 ksec, then the probability of detecting only 0 or 1 photon is only 1.7\%.  We therefore take 6 counts as a reasonable, high-confidence, upper limit.  This corresponds to a count rate, $\mathcal{R}_x$ of:
\begin{equation}
\mathcal{R}_x \lesssim 6/40300 = 0.00015 ~{\rm counts~s}^{-1}  $$
\end{equation}
{\bon We then use the {\tt PIMMS} software (Mukai 1993) to estimate the limits on the unabsorbed X-ray flux corresponding to $\mathcal{R}_x$. We estimated the unabsorbed flux in two energy ranges: 0.5-6 keV, which is fully covered by ACIS-S data and hence relatively insensitive to the assumed spectral shape of the object; and 0.1-100 keV, effectively bolometric, which required extrapolation outside the ACIS-S energy range, and hence sensitive to the assumed spectrum.}

We focus on single temperature plasma emission computed with the {\tt APEC} code (Smith et al.~2001) for different temperatures and an assumed absorbing column of $N_H=3 \times 10^{20}\,{\rm cm}^{-2}$.  At the highest temperatures, plasma emission is dominated by the bremsstrahlung continuum, and the {\tt APEC} model is only available up to $kT=64$ keV; we have therefore also included bremsstrahlung model spectra up to $kT=80$ keV. These results are summarised in Table \ref{tbl:xray}.

As discussed in the following sections, the {\em APEC} plasma model is the one we prefer.  However, since the spectral shape of the emission at these low accretion rates is quite uncertain, we also show in Table \ref{tbl:xray} the conversion of our upper limit on the count rate to X-ray flux limit using several other typical model spectra.  {\bon As suggested above and} can be seen from the table, the derived limits on the flux are not particularly sensitive to details as long as much of the emission is contained within the {\em Chandra} band ($\sim$0.5-6 keV) and is not severely absorbed.  This point is discussed further in Sect.~\ref{sec:otherbands}.

Finally, for an adopted source distance of $175 \pm 25$ pc we can convert the X-ray flux limits to constraints on the X-ray luminosity.  The average of the log $L_x$ values in Table \ref{tbl:xray} for different spectral parameters leads to a limit on the {\bon 0.1-100 keV} X-ray luminosity of
$$ L_x \lesssim 2 \times 10^{28} ~{\rm ergs~s}^{-1} $$
which we discuss in more detail below.  

{\bon In the following two sections (\ref{sec:Mdotlimit} and \ref{sec:otherbands}), we attempt to interpret how the limit on $L_x$ informs us about the mass accretion rate, $\dot M$, onto the white dwarf. This is by far the most difficult of the steps needed to constrain $\dot M$. The relation between $\dot M$ and $L_x$ depends on such factors as the importance of the white dwarf's magnetic field, what fraction of the surface of the white dwarf experiences accretion, the emission mechanisms, and so forth.  These are explored in Sect.~\ref{sec:otherbands} which enables us to estimate the limits on $\dot M$ given in the last column of Table \ref{tbl:xray}.}

\subsection{Limits on Mass Accretion Rates}
\label{sec:Mdotlimit}

The photosphere of WD 1145+017, with an effective temperature of 15,900 K (V15) is far too cold to produce a detectable level of X-ray emission in the absence of accretion. However, X-ray emission can be expected from WD 1145+017 because it is likely accreting the sublimated gaseous remains\footnote{We note, however, that in principle, larger objects, e.g., $\gtrsim 100 \,\mu$m, may resist sublimation and accrete directly onto the surface of the white dwarf, thereby depositing accretion energy directly into the surface layers.} of the dust that we infer from the dips in optical flux.  Accreting white dwarfs produce thermal, optically-thin, collisionally-excited plasma emission above the surface and/or thermal, optically-thick, blackbody-like emission from the heated photosphere (see Mukai 2017 for a review). Here we rely heavily on the well-studied accreting white-dwarf systems, mainly cataclysmic variables (CVs), extrapolating to lower accretion rates from the systems with the lowest known accretion rates (see Sect.~\ref{sec:otherbands}).

The X-ray properties of an accreting white dwarf differ markedly depending on whether it has a magnetic field strong enough to control the accretion flow. There are high-resolution spectroscopic observations of WD 1145+017 (Xu et al.~2016) that did not exhibit any Zeeman splitting, yielding an upper limit on the magnetic field of $\sim$10 kG (P.\,Dufour \& F.\,Hardy private communication). This is a factor of $\sim$1000 lower than typically found in magnetic CVs.   To estimate the magnetic channeling of the ion flow, we utilize the well-known dimensional-analysis result for the magnetospheric radius of a rotating magnetised white dwarf with an orbiting collection of charged atoms:
\begin{equation}
R_m \simeq \zeta^{2/7} (GM_{\rm wd})^{-1/7} \dot M_{\rm acc}^{-2/7} B_{\rm wd}^{4/7} R_{\rm wd}^{12/7}
\label{eqn:Rm1}
\end{equation}
(see, e.g., Pringle \& Rees 1972; Lamb et al.~1973; Klu\'zniak \& Rappaport 2007) where $M_{\rm wd}$, $R_{\rm wd}$, and $B_{\rm wd}$ are the mass, radius, and surface magnetic field of the white dwarf, $\dot M_{\rm}$ is the accretion rate, and $\zeta$ is a dimensionless factor that depends on such quantities as the ratio of radial velocity to orbital velocity and solid angle subtended by the accretion flow.  Plugging in illustrative parameter values for WD 1145+017, we find:
\begin{equation}
R_m \simeq 13 \, R_{\rm wd} \left(\frac{\zeta}{0.01}\right)^{2/7} \left(\frac{\dot M_{\rm acc}}{10^{11}\, {\rm g/s}}\right)^{-2/7} \left(\frac{B_{\rm wd}}{10 \,\rm kG}\right)^{4/7} 
\label{eqn:Rm2}
\end{equation}
By comparison, the corotation radius is 
\begin{equation}
R_c \simeq \left(\frac{GM_{\rm wd}}{4 \pi^2}\right)^{1/3} P^{2/3} \simeq 95 \, R_{\rm wd} \left(\frac{P}{4.5 \, {\rm hr}}\right)^{2/3}
\end{equation}
where $P$ is the rotation period of the magnetic white dwarf normalised for convenience to the typical dip period, though this may well be, or not, the rotation period of the white dwarf.  For motivation on why this choice for the rotation period might be of interest, see the recent model proposed by Farihi et al.~(2017).

From the above estimates, we can see that the magnetosphere would lie within the corotation radius for a 4.5 hour period\footnote{We reiterate that the rotation period of WD 1145+017 is unknown.  However, in the Farihi et al.~(2017) model, the rotation period of the white dwarf must be 4.5 hours.} unless the magnetic field is $\gtrsim 300$ kG or if $\dot M_{\rm acc} \lesssim 10^8$ g s$^{-1}$.  Since neither of these seems likely, especially during the `high activity' level of this past observing season, we take the magnetospheric radius to lie somewhat inside the corotation radius corresponding to 4.5 hours.  If the white dwarf is indeed rotating with a 4.5 hour period, or longer, this would be considered a ``slow-rotator'' in the language of accreting neutron stars (see, e.g., Klu\'zniak \& Rappaport 2007), which is capable of accreting any charged atoms reaching the magnetosphere.  Finally, in this regard, we note for $B_{\rm wd} \lesssim 100$ G, the ion magnetosphere would lie inside the white dwarf.

Because we have no firm estimate of either the white dwarf rotation period or its B field, we therefore proceed by assuming that WD 1145+017 can accrete either via a magnetically controlled accretion flow or via an accretion disk.  Any accreting matter originating from a 4.5 hr orbit will retain either all, or half, of the kinetic energy it acquired from the gravitational potential, respectively, for the two cases.  The maximum possible X-ray luminosity that is available from the accretion luminosity is therefore:
\begin{equation}
L_x \simeq \frac{G M_{\rm wd} \dot M_{\rm acc}}{R_{\rm wd}}
\label{eqn:mdot}
\end{equation}
The upper limit on X-ray luminosity ($2 \times 10^{28}$ ergs s$^{-1}$; see Table \ref{tbl:xray}) then corresponds to $\dot M < 2 \times10^{11}$ g s$^{-1}$ where we have taken the mass and radius of WD 1145+017 to be 0.6 $M_\odot$ and 1.4 $R_\oplus$, respectively.

An important caveat, however, is that this limit holds only if the bulk of the released gravitational potential energy emerges in the X-ray band of the spectrum.  At these low accretion rates onto a white dwarf, this is a complicated issue. We next tackle the question of when, and under what conditions, the gravitational energy release comes out predominantly in the X-ray band.

\subsection{In What Energy Band Does the Accretion Luminosity Appear?}
\label{sec:otherbands}

In the magnetic case, the flow inside the magnetospheric radius is confined by the magnetic field and accretion will occur on a relatively small region (`spot') near the magnetic poles. The fractional surface area of the spot, $f$, is as important in determining the outgoing radiation as is the total accretion rate, $\dot M_{\rm acc}$.  In more recent studies, the parameter of choice is often the specific accretion rate (accretion rate per unit area), $\dot m_{\rm acc}$, instead of $\dot M_{\rm acc}$.

The framework for interpretation in the case of magnetic CVs is the Aizu model (Aizu 1973), in which a free-falling (supersonic) flow hits the white dwarf surface and forms a standing shock with a shock temperature, $T_s$, that is determined by the free-fall velocity of the flow 
\begin{eqnarray}
T_s & = & \frac{3}{8} \frac{GM_{\rm wd} m_p\mu}{R_s} \nonumber \\
 & \simeq  & 68 \left(\frac{M_{\rm wd}}{0.6 \, M_\odot}\right) \left(\frac{1.4\, R_\oplus}{R_s}\right) \left(\frac{\mu}{2}\right) ~{\rm keV}
\end{eqnarray}
(see, e.g., Frank et al.~1985) where $R_s$ is the radial distance of the shock front (presumed to be near the white dwarf), $m_p$ is the proton mass, and $\mu$ is the dimensionless mass per particle in the post-shock region.  We take the post-shock material to be essentially fully ionized metals for which $\mu \simeq 2$. While any contribution from H would lower $\mu$ somewhat, it is probably nowhere as low as the typically assumed value of $\sim$0.6 which is appropriate for solar composition material.  The post-shock plasma cools via bremsstrahlung emission and the shock height is determined by the requirement that the plasma cooling time equals the remaining infall time from the shock to the surface.

In magnetic CVs, $f$ is variously estimated to have values in the 10$^{-4}$ to 10$^{-2}$ range, with $\dot m_{\rm acc}$ of order 1 g s$^{-1}$ cm$^{-2}$, which places the shock close to the white dwarf surface.  In this case, roughly half the X-rays are absorbed by the white dwarf photosphere, and the remaining half provides the observed luminosity.

Unless $f$ is much smaller in WD 1145+017 than in magnetic CVs, the specific accretion rate is likely very low ($\lesssim 10^{-4}$ g cm$^{-2}$ s$^{-1}$). This should result in a large shock height and a post-shock plasma of much lower density than in magnetic CVs. In that case, only a negligible fraction of emitted X-rays would be intercepted by the white dwarf surface. On the other hand, we must consider the possibility that X-rays are not emitted at all.

In principle, the post-shock plasma in WD 1145+017 could cool via cyclotron emission; however, it is likely to be in the bremsstrhalung-dominated regime (Masters et al.~1977). If accretion takes place predominantly in the form of dense blobs (Kuijpers \& Pringle 1982), then it is possible to avoid the formation of a shock above the photosphere, instead
liberating all the potential energy below it. However, given the very low total accretion rate, this scenario is unlikely unless strong density contrasts from individual grains, which produced the gas, can survive. The large shock height itself will lower the shock temperature---in the extreme case, when the shock height approaches the magnetospheric radius, this will be a factor of $\sim$10 (i.e., 0.13 $R_\oplus/1.4 \, R_\oplus$) lower, resulting in an X-ray temperature of $\sim$7 keV.  

It is possible that additional effects come into play that we have not considered (e.g., how quickly can ions and electrons reach temperature equilibrium, and what are the consequences of such a two-temperature plasma?).

In the case of a Keplerian flow (i.e., in the non-magnetic case), the material will encounter a strong shock just above the white dwarf surface.  Its temperature is then determined by the Keplerian velocity, or equivalently, the depth of the gravitational potential, and is $T_s \simeq 35$ keV for a 0.6 $M_\odot$ white dwarf. At low accretion rates, the shocked gas is optically thin, and cools by emitting thermal X-rays (including atomic X-ray line emission) before settling onto the white dwarf. Therefore the average temperature of X-ray emitting gas is of order half the shock temperature.  

Dwarf novae in quiescence are non-magnetic CVs with low accretion rate, and their X-ray data generally agree with this simple expectation (Byckling et al.~2010), with a typical luminosity of $10^{30}$--$10^{32}$ erg s$^{-1}$.  However, at the lowest accretion rates, the shock appears to be at a considerable height above the surface, and hence at a lower temperature. GW Lib is the best-studied case with a characteristic temperature of 2 keV and and an X-ray luminosity of $\sim$$1 \times 10^{29}$ erg s$^{-1}$ (Hilton et al.~2007). This suggests that the shock is located several white dwarf radii above the surface. Our count rate-to-flux conversion for WD 1145+017 (see Table \ref{tbl:xray}) allows for temperatures that are even this low.

As discussed above, for either magnetic or disk accretion the shock temperatures encountered in WD 1145+017 are plausibly confined to a range of 8-70 keV, and in the optically thin case (which likely obtains), the mean X-ray emitting gas should have something of the order of half these temperatures.  As we can see from Table \ref{tbl:xray}, the conversion factors from detected X-ray counts to bolometric luminosity cover only a factor of $\sim$4 in range. Therefore, if we adopt a mean effective X-ray emitting temperature of $kT_{\rm emis} \simeq 20$ keV, we can set an upper limit on the X-ray luminosity of $L_x \lesssim 1.7^{+1.7}_{-0.8} \times 10^{28}$ ergs s$^{-1}$ and a corresponding limit on the mass accretion rate of $\dot M \lesssim 2^{+2}_{-1} \times 10^{11}$ g s$^{-1}$.  The ranges on the limits reflect the uncertainty in the spectral models.

One potential complication that we have not addressed is the absorption of X-rays by circumstellar matter, the latter of which is observed in the {\em optical} (Xu et al.~2016). In a medium with solar abundance, the most important elements for X-ray absorption are C, N, O, and Fe, and a concomitant hydrogen column density of order $10^{22}$ cm$^{-2}$ is sufficient to alter the ACIS-S count rate-to-flux conversion factor by 2. Such a total column density will have an Fe column density of $2 \times 10^{17}$ cm$^{-2}$. Xu et al.~(2016) report that the FeII line has a column density of $10^{16}$ cm$^{-2}$, so it is quite reasonable to expect that the total metal column density will be $\lesssim 10^{17}$ cm$^{-2}$, and therefore not substantially alter our estimates above.

Finally, we show that even if a substantial accretion rate of unsublimated matter rains down onto the white dwarf surface, the resultant thermal emission will always be very far below the {\em Chandra} X-ray band.  The thermalised surface temperature due to direct accretion over a solid angle $\Delta \Omega$ onto the surface of the white dwarf is given by
\begin{equation}
T_{\rm wd, ther} \simeq \left[\frac{G M_{\rm wd} \dot M}{\Delta \Omega \sigma R_{\rm wd}^3} \right]^{1/4} \simeq 3800 \left(\frac{1\,{\rm sr}}{\Delta \Omega}\right)^{1/4}\,\dot M_{11}^{1/4} ~{\rm K}
\end{equation}
where $\sigma$ is the Stefan-Boltzmann constant and $\dot M_{11}$ is the accretion rate in units of $10^{11}$ g s$^{-1}$.  Since this temperature is lower than that of the white dwarf itself, the resultant emission is nowhere near the X-ray band.

\begin{table*}
\centering
\caption{Observed Phenomena vs.~Scenarios}
\begin{tabular}{llll}
\hline
\hline
Phenomenon      &       Dust Emitting Asteroids$^a$          &               Asteroid Collisions $^b$              &                    Magnetic Shepherding$^c$ \\
\hline
Long-lived dips$^d$     &       Straightforward to explain       &          Difficult to explain the persistence             &          Tricky to explain but plausible \\
Multiple periods$^e$     &      Straightforward to explain        &          Straightforward to explain                      &             Difficult to explain \\
Depth of dips$^f$         &      Difficult to explain the depth      &         Straightforward to explain                  &                 Straightforward to explain \\
Period stability$^g$      &       Partially straightforward          &           Depends on numbers, masses of bodies                             &                                  Straightforward for one period \\
Sudden events$^h$       &     Plausible                     &                       Natural explanation                             &                                  Neutral \\
Activity changes$^i$     &     Plausible                      &                      Natural explanation                               &                                Neutral \\
Source of dust      &       Natural explanation     &                      Natural explanation                                 &                 No explanation \\
\hline
\hline
\end{tabular}
\label{tbl:phenomena}

\scriptsize{ (a) Emission via sublimation, rotational instability, or thermal fracture; see also Vanderburg et al.~(2015).  (b) Kenyon \& Bromely (2017).  (c) Farihi et al.~(2017). (d) Weeks to months.  (e) They range over 8\% in period.  (f) Up to 60\%.  (g) See discussion below.  (h) E.g., event `G6420' in (G17).  (i) See figure \ref{fig:activity}.} 

\end{table*}

\subsection{Comparison to Inferred $\dot M$ Based on Dust}
\label{sec:dustaccr}

Inferring the rate of mass accretion onto the white dwarf WD 1145+017 is at least a few steps removed from the observations of periodic dips in the optical flux.  The first step is to estimate the amount of dust present at any given moment, and from that the rate of mass loss from the orbiting bodies in the form of dust. This of course does not factor in any gas that is lost along with the dust.  Once the dust is in orbit about the white dwarf it has an uncertain lifetime against sublimation.  From there, even assuming that the gaseous material ultimately makes its way down to the surface of the white dwarf, we do not know the time lag between its production and its accretion.  Finally, since we know that the dust production changes dramatically on a year timescale (see Fig.~\ref{fig:activity}), we cannot even treat this set of steps as a steady-state process.  

Nonetheless, we will proceed to get a sense of what the accretion rate might be.  Vanderburg et al.~(2015) estimated a dust production rate in WD 1145+017 of $2 \times 10^9 - 3 \times 10^{10}$ g s$^{-1}$.  The dust activity level during the 2016-2017 season was, however, at least 40 times higher than in 2015 when it was first discovered.  Thus, the dust mass loss rate could well lie in the range of $10^{11}-10^{12}$ g s$^{-1}$ at the current epoch.

Here we briefly review the arguments by which we infer the dust mass loss rate.  During the interval over which the {\em Chandra} observations were conducted, the mean flux depression over an orbital cycle is close to 15\%.  If the optimum dust grain cross section per unit mass occurs near 1 $\mu$m, then we can consider a sheet of dust that is 1 $\mu$m thick, covers the white dwarf's diameter, and is 15\% filled.  If such a sheet runs around the entire orbit, the amount of mass it would contain is $\Delta M \simeq 3 \times 10^{16}$ g.

A key unknown parameter is the lifetime of this dust.  For dust grains larger than $\sim$2 $\mu$m, which may survive for weeks without sublimating (see Xu et al.~2017), we can still estimate an upper limit to the grain lifetime in the following way.  We tentatively assume that the dust emitted by the orbiting debris occupies a volume that is roughly the same size in the vertical (to the orbital plane) direction as in the radial direction.  In that case, the dust clouds extend at least a white dwarf radius in the radial direction, or $\sim$1\% of the orbital radius.  This would tend to lead to complete shearing of the unbound material around the entire orbit within about 60 orbits, or about 10 days\footnote{This hinges on there being no significant B field for WD 1145+017; see however Farihi et al.~(2017) for an interesting scenario involving a substantially magnetised white dwarf.}.  In fact, in order to maintain coherent dip structures as are seen in Figs.~\ref{fig:LCs} and \ref{fig:superposed}, that are as narrow as 0.1 in orbital phase, we can infer that the dust replenishment time must likely be as short as a day.  We postulate, therefore, that the dust that we see at any given moment, must be replenished every day or so at a minimum mass loss rate of $\sim$$2 \times 10^{11}$ g s$^{-1}$.  This is quite consistent with the ballpark range mentioned above, based on a simple `activity' scaling from the K2 discovery epoch.  

This estimate of the lower limit on the rate of mass (from dust) that must ultimately be accreted by the white dwarf is just about the same as is the upper limit inferred from our X-ray observations.  Therefore, the best we can say at the moment is that the X-ray {\em upper limit} on $\dot M$ is closely consistent with the {\em lower limit} on $\dot M$ that is inferred from dust sublimation rates which come from the observed dips in the optical flux.

\section{Discussion}
\label{sec:discuss}

In this work we have assumed that the dips in photometric flux from WD 1145+017 are caused by dust clouds circling around the white dwarf.  It would seem that dust is the only efficient form of matter capable of blocking up to 60\% of system light for intervals of time longer than a major planet would take to transit the white dwarf.  We presume that the origin of the dust is ultimately from collisions or sublimation of small bodies, i.e., asteroids, orbiting the white dwarf.  The passage of the dust clouds has characteristic periods of 4.5-4.9 hours.  In our judgement, these likely reflect the orbital periods of the bodies which give rise to the dust.  However, recently an alternative idea has been put forth wherein the characteristic periodicity represents the rotation period of the white dwarf which is presumed to be strongly magnetised.  The concentration of dust at certain rotational longitudes would then be explained by magnetic shepherding of charged dust grains of submicron size (Farihi et al.~2017).

We now proceed to list some of the empirical facts that we have learned about the photometric dips in WD 1145+017 from this work as well as from previous studies.  A number of the dip features and properties can be thought about in the context of (i) orbiting asteroids; (ii) sublimation of, and collisions among, the asteroids to produce dust clouds; and (iii) possibly magnetic shepherding of charged dust grains by an hypothesised strongly magnetised white dwarf.  At present, there are no quantitative models that can definitively relate any of these models to specific details of the observations.  Nonetheless we will comment about the possible applicability, or lack thereof, of some of the basic theoretical ideas with the observational facts, as we see them.  

Some of the basic phenomena that we are trying to explain, along with different physical models, are summarised in Table \ref{tbl:phenomena}.  We discuss these in more detail below.

Dip features are persistent from orbit to orbit (see e.g., Figs.~\ref{fig:superposed} and \ref{fig:waterfall}). Some of the features evolve over days to weeks, but a few are present for at least 8 months (see Figs.~\ref{fig:waterfall} and \ref{fig:WFprevious}).  The ``A'' period, and other periods within $\pm 0.1\%$ of it, seem to be almost always detectable from ground-based observations (V15; G17; G\"ansicke et al.~2016; R16).  The ``B'' period has been detected over only one two-week period from the ground (G17). The ``C-F'' periods were seen only with {\em Kepler} K2, presumably due to their shallow depths.  Some large, i.e., deep and broad, dip features (perhaps actually combinations of independent dip features) can last for up to 30\% of an orbital cycle (e.g., Figs.~\ref{fig:LCs}, \ref{fig:superposed}, and \ref{fig:waterfall}).  Because these broad dips largely repeat from orbit to orbit, and can even persist for up to several weeks (see Fig.~\ref{fig:superposed}), it is difficult to know whether they are due to a single complex dust cloud or a collection of independent dust clouds.  It is also difficult to distinguish between dips simply evolving in shape or being comprised of a set of dips with slightly different orbital periods. 

The most persistent dips produce lines on a waterfall plot that are consistent with being straight (e.g., Feature 1 in Fig.~\ref{fig:waterfall} and the unlabelled long vertical stripe in Fig.~\ref{fig:WFprevious}), i.e., representing a period that is constant to within 2 parts in $10^5$.  However, there are some indications of period changes on timescales of $P/\dot P$ of $\lesssim 2000$ years (Feature 1 in Fig.~\ref{fig:waterfall}).  Feature 2 in Fig.~\ref{fig:waterfall} appears to have a continuously changing period with $P/\dot P \simeq 400$ years.  For the larger and more complex dips (Fig.~\ref{fig:superposed} and Features 4, 5, and 6 in Fig.~\ref{fig:waterfall}), we can envision a collection of small asteroids in a near common orbit, each of which produces a dust cloud.  The dust clouds can overlap.  As the small bodies evolve in size and position, their dust clouds drift slowly with respect to each other.  

Due to the fact that any dust clouds capable of producing the observed dips must necessarily be far outside any asteroid's Hill sphere, there must necessarily be large shearing effects on the dust clouds.  In the absence of any other shepherding forces, we expect that the dust clouds would not last for more than a couple of days before dissipating.  If, in fact, the dust clouds quickly dissipate due to shearing, we would expect that the dust must be continually replenished on this timescale. However, in the scenario recently suggested by Farihi et al.~(2017), if the dust grains are small ($\lesssim 0.05 \,\mu$m) and charged (due to UV bombardment), and the white dwarf has a surface magnetic field as large as $\sim$100 kG\footnote{The magnetic field in WD 1145+017 has not yet been measured well enough to know whether the magnetic model is viable, though recent indications are that $B \lesssim 10$ kG (Sect.~\ref{sec:Mdotlimit}).  Moreover, recent measurements of dust grain size by Xu et al.~(2017) seem to indicate that very few, if any, grains less than 1 $\mu$m are likely to remain in the dust clouds.}, the dust can be forced to corotate with the star via the magnetic field.   In that case, one would interpret the periodicity of the dust dips as being due to the underlying rotation period of the white dwarf rather than asteroid orbital periods.

Empirically, the dust clouds are able to obscure up to $\sim$60\% of the light of the white dwarf, but they have never been observed to block more than this apparent upper limit of 60\%.  It remains a mystery as to whether this indicates that the dust covers the entire star with a minimum transmission of 40\%, or if the optical depth can become very high but never cover more than 60\% of the geometrical projected area of the white dwarf.  In order to block any substantial fraction of the white dwarf's light, the dust cloud must extend away from the orbital plane by of order 1/2$^\circ$ (as seen from the WD).  Assuming that dust in at least some deep dips arises from individual asteroids traveling at $\sim$300 km s$^{-1}$, we conclude that the dust would have to be ejected with speeds of $\sim$3 km s$^{-1}$ or else the relatively large range of angular tilts could not be accommodated dynamically. Thermal velocities, i.e., involved in sublimation, would be inadequate to propel material far enough above the orbital plane.  However, occasional collisions among orbiting asteroids could yield high-velocity fragments.  Another explanation could involve the magnetic shepherding model of Farihi et al.~(2017) where small charged dust grains could fall in toward the white dwarf while being channeled by the magnetic field onto its poles.  In the process, the grains could then naturally reach high latitudes on the white dwarf.

Eventually, the dust grains must sublimate or be dragged in toward the white dwarf via Poynting-Robertson effect.  In the Farihi et al.~(2017) scenario, the small charged dust grains would naturally fall toward the white dwarf surface as they cross the magnetosphere and are swept up by the magnetic field\footnote{This assumes that the corotation radius lies outside the magnetosphere}.  Presumably, the dust grains would sublimate before they ever hit the white dwarf surface.  Obviously, the details of all these processes, as they operate in WD 1145+017, are highly uncertain.  However, some of the gas that results from sublimation becomes ionised and finds itself moving at high speed around and toward the white dwarf.  We can speculate that it is this high-velocity gas that may be responsible for the broad absorption line features seen in high-resolution spectral observations (Xu et al.~2016; Redfield et al.~2017).

As we have shown in Sect.~\ref{sec:Mdotlimit}, if WD 1145+017 has a surface B field above $\sim$100 G, the magnetosphere for gas ions would lie outside the white dwarf (see Eqn.~\ref{eqn:Rm2}).  Once the ions cross the magnetosphere, they will be forced to corotate with the white dwarf, whatever its rotation period may be (perhaps even 4.5 hours; Farihi et al.~2017).  The inferred rate of gas release from sublimating dust grains could be as high as $10^{12}$ g s$^{-1}$ (see Sect.~\ref{sec:dustaccr}) during the most active intervals of WD 1145+017.  The upper limit on the X-ray luminosity of $L_x \lesssim 2 \times 10^{28}$ ergs s$^{-1}$ is marginally consistent with the estimated lower limit on $\dot M$ inferred from the dust transits.  As a guide to future X-ray observations, we note that the accretion of the gas ions is highly likely to take place via magnetic funnelling.

\section{Summary and Conclusions}
\label{sec:conclusion}

In this work we have chronicled the photometric behaviour of WD 1145+017 during an 8-month period comprising the 2016-2017 observing season.  The overall dipping activity (Fig.~\ref{fig:activity}) was the highest that it has been since the source's discovery three years ago (V15).  We present a number of orbital-cycle-long lightcurves to illustrate the behaviour of the dips during the season, and we also show how the dips evolve slowly over a couple of two-week intervals.  A new type of `waterfall diagram', is presented which gives an overview of dipping behaviour over the entire season.  The major dipping features all had a period of 4.4913 hours, or within 0.1\% of that period. There was no sign of the K2 periods `B' through `F', though the `B' period was detected from our ground-based observations of the previous year.  

We also carried out four 10-ksec {\em Chandra} observations of this source, three of which were done during the optical monitoring campaign.  No significant X-ray flux was detected during the 40 ksec total exposure (see Fig.~\ref{fig:chandra} and Table \ref{tbl:xray}).  We use this to set a secure upper limit on the average X-ray flux of $2 \times 10^{-15}$ ergs cm$^{-2}$ s$^{-1}$ (0.5-6 keV band).  This translates to an upper limit on the bolometric X-ray luminosity during this period of $2 \times 10^{28}$ ergs s$^{-1}$.  If all the accretion luminosity came out in the X-ray band, this would correspond to an upper limit of $\dot M_{\rm acc} \lesssim 2 \times 10^{11}$ g s$^{-1}$.  This is comparable with the inferred rates of dust production in the system, and, by inference, the rate of that sublimated material onto the surface of the white dwarf.  However, we caution that the nature of the accretion-induced X-ray spectrum at these low values of $\dot M$ is not well understood (see Sect.~\ref{sec:otherbands}), and a significant fraction of the accretion energy may not come out in the X-ray band.  

We do not yet have a good picture of the nature of the bodies orbiting WD 1145+017.  However, it is clear that the dust clouds change on timescales of weeks to years, and persistent monitoring may yet reveal the number of bodies producing the dust clouds and a better estimate of their masses.

Data files for all of the ground-based photometric observations presented  this work are available upon request from author BLG. These files include normalised fluxes for the lightcurves, as well as their dip solutions (BJD, depth, ingress time, egress time). Web page URLs are also available from BLG that show all lightcurves in detail and in several formats.

\vspace{0.3cm}
\noindent
{\bf Acknowledgements}

We thank Paul Benni and Tom Kaye for providing a few of the lightcurves and for helpful discussions.  S.\,R.~and B.\,L.\,G.~acknowledge partial support from NASA {\em Chandra} grant GO7-18008X.  A.\,V.~was supported by the NSF Graduate Research Fellowship, grant No.~DGE 1144152, and also acknowledges partial support from NASA's TESS mission under a subaward from the Massachusetts Institute of Technology to the Smithsonian Astrophysical Observatory, Sponsor Contract Number 5710003554.  This work was performed in part under contract with the California Institute of Technology (Caltech)/Jet Propulsion Laboratory (JPL) funded by NASA through the Sagan Fellowship Program executed by the NASA Exoplanet Science Institute.

%%%%%%%%%%%%%%%%%%%%%%%%%%%%%%%%%%%%%%%%%%%%%%%%%%



%\bibliographystyle{mnras}


%%%%%%%%%%%%%%%%%%%%%%%%%%%%%%%%%%%%%%%%%%%%%%%%%%

%%%%%%%%%%%%%%%%% APPENDICES %%%%%%%%%%%%%%%%%%%%%

%\appendix

%\section{Any Appendices Go Here}
%\label{app:1}

%%%%%%%%%%%%%%%%%%%%%%%%%%%%%%%%%%%%%%%

% Don't change these lines
\bsp	% typesetting comment
\label{lastpage}
\end{document}